\documentclass[aps,prb,floatfix,reprint]{revtex4-1}
\usepackage[T1]{fontenc}
\usepackage[latin9]{inputenc}
\usepackage[a4paper]{geometry}
\usepackage{xcolor}
\usepackage{verbatim}
\usepackage{textcomp}
\usepackage{amsmath}
\usepackage{amssymb}
\usepackage{stackrel}
\usepackage{graphicx}
\usepackage{wasysym}
\usepackage{hyperref}
\usepackage{xcolor,soul}

\begin{document}
\title{
Current-induced forces in single-resonance systems.
}
\author{Sebasti\'an E. Deghi,$^1$ Lucas J. Fern\'andez-Alc\'azar,$^2$ Horacio
M. Pastawski,$^1$ and Ra\'ul A. Bustos-Mar\'un$^{1,3}$}

\affiliation{$^1$Instituto de F\'{\i}sica Enrique Gaviola and Facultad de
Matem\'{a}tica, Astronom\'{\i}a, F\'{\i}sica y Computación, Universidad Nacional de
C\'{o}rdoba, Ciudad Universitaria, C\'{o}rdoba, 5000, Argentina\\
$^2$Wave Transport in Complex Systems Lab, Department of Physics, Wesleyan University, 
Middletown, CT-06459, USA\\
$^3$Facultad de Ciencias Qu\'{\i}micas, Universidad Nacional de C\'{o}rdoba, Ciudad
Universitaria, C\'{o}rdoba, 5000, Argentina}

\begin{abstract}
In recent years, there has been an increasing interest in nanoelectromechanical devices, 
current-driven quantum machines, and the mechanical effects of electric currents on nanoscale conductors.
Here, we carry out a thorough study of the current-induced forces and the electronic friction of systems whose electronic effective Hamiltonian can be described by an archetypal model, a single energy level coupled to two reservoirs.
Our results can help better understand the general conditions that maximize the performance of different devices modeled as a quantum dot coupled to two electronic reservoirs.
Additionally, they can be useful to rationalize the role of current-induced forces in the mechanical deformation of one-dimensional conductors.
\end{abstract}
\maketitle

\section{INTRODUCTION}

As it is nowadays clear, electric currents can induce mechanical forces in nanodevices.
Such current-induced forces (CIFs), sometimes dubbed ``electron wind forces'', have attracted increasing attention in many areas of Condensed Matter physics, including molecular electronics,
\cite{chiaravalloti2007,dundas2009,bode2011,kudernac2011,tierney2011,bustos2013,
kim2014,cunningham2014,fernandez2015,lu2015,ludovico2016,lu2016,gu2016,celestino2016,
calvo2017,fernandez2017,hopjan2018,ludovico2018,fernandez2019,chen2019,lin2019,mcvooey2020} nanoelectromechanics,\cite{naik2006,stettenheim2010} electromigration,\cite{hoffmannvogel2017,chatterjee2018} and quantum thermodynamics.\cite{ludovico2016entropy,benenti2017,whitney2018,bustos2019,deghi2020,zimbovskaya2020}
Among the most interesting examples are those where the forces are nonconservative.
These types of forces may result in phenomena such as cooling, heating, or amplification of mechanical motions.~\cite{naik2006,stettenheim2010,lu2015,lu2016} %
Furthermore, it is also the basis of exciting new proposals such as adiabatic quantum motors.~\cite{bustos2013}
The most simplified description of a quantum motor consists of a system connected to leads and where a current of quantum particles drives some mechanical degrees of freedom.
This is the reverse of what happens in a quantum pump, where the driving Hamiltonian's parameters, which may arise from a mechanical or an external electrical manipulation, induces a current. \cite{brouwer1998,bode2011,bustos2013,fernandez2017,ludovico2016entropy,whitney2018,bustos2019}
Understanding the rules that dictate the interplay between the electronic and the mechanical degrees of freedom is crucial for the design of optimal electromechanical devices, and it can even contribute to the emergence of devices with novel features.

Aside from the potential applications, the type of devices discussed above may result particularly attractive from the theoretical point of view.
This is so as quantum motors, quantum pumps, and some nanoelectromechanical devices can be interpreted as macroscopic manifestations of quantum behavior. Thus, these systems may provide new insights into the transition between the classical and quantum worlds, both from a dynamical
~\cite{fernandez2015,fernandez2017} and a thermodynamical
~\cite{zimbovskaya2020,whitney2018,bustos2019,deghi2020}
perspective.
Moreover, the effective Hamiltonian of open quantum systems turns out to be non-Hermitian, which adds an extra richness to the problem.
For example, the non-Hermiticity makes the dynamics of electrons to be affected in a nontrivial way by the 
change of some parameters, especially close to the so-called quantum dynamical phase transitions (QDPTs). \cite{pastawski2007,dente2008,rotter2010,bustos2010,garmon2012,rotter2015,ruderman2016,sanchez2017,gorbatsevich2017,garmon2019}
\begin{figure}[ht]
\begin{centering}
\includegraphics[width=2.5in]{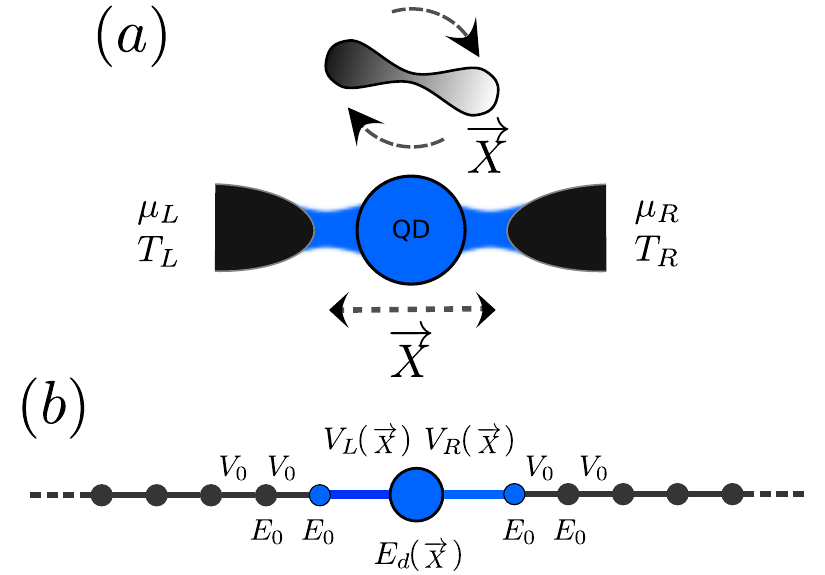}
\end{centering}
\caption{\textbf{(a) -} Schematic representation of a quantum dot coupled to two electronic reservoirs at different chemical potentials ($\mu_L$ and $\mu_R$) and temperatures ($T_L$ and $T_R$).
The quantum dot also interacts with some mechanical degrees of freedoms ($\vec{X}$) that change the energy of the dot or its coupling to the reservoirs.
The mechanical degrees of freedom may include a motion of the quantum dot itself or the motion of some device capacitively coupled to it.
\textbf{(b) -} Tight-binding model of the electronic Hamiltonian, see Sec.\ref{sec:TB}.
Site energies and hoppings are denoted by $E$ and $V$, respectively. In blue, we highlight the sites that correspond to the system in our model.
}
\label{fig:1}
\end{figure}

In this work, we perform an in-depth study of the current-induced-forces (CIFs) and the electronic-friction tensor of devices where a single energy level is relevant and, hence, they can be described by one of the most common Hamiltonian models in quantum transport,~\cite{haug2008,pastawski2001,zimbovskaya2013} see Fig. 1.
The main motivation behind that is to find the general conditions that lead to the improvement of the performance of different forms of nanoelectromechanical devices and quantum machines.
Moreover, since CIFs may lead to the mechanical failure of conducting devices such as nanowires, the present work may also contribute to a better understanding of the circumstances that increase the chances of such failures.

The work is organized as follows. In Sec. \ref{sec:general_theory} we discuss the main theoretical aspects of the studied system. These include, a brief introduction to the QDPTs and current-induced forces relevant to our problem, Secs. \ref{sec:DPT} and \ref{sec:CIFs} respectively. In Sec. \ref{sec:results} we present the results of this work and in Sec. \ref{sec:conclusions} we highlight the main findings and discuss their importance.

To avoid overwhelming the readers with the derivations of the many analytic expressions, we put all the nonessential comments and mathematical details in the appendices from \ref{app:U^eq} to \ref{app:friction_TB}.
  
\section{General theory.\label{sec:general_theory}}

As we will see, the understanding of CIFs and electronic friction requires the comprehension of the different quantum dynamical regimes of the electrons.
These regimes are separated by QDPTs, abrupt change in the decay dynamics of particles. 
As we will also see, this is intimately related to the maximization of the work done by CIFs.
For that reason, in the first part of this section, we briefly review the QDPTs that our model Hamiltonian may undergo, before going to the theoretical description of CIFs.

\subsection{Quantum Dynamical Phase Transitions.\label{sec:DPT}}

The intuitive idea of dynamical phase transitions is better understood by considering the classical problem of a damped harmonic oscillator in the absence of external forces.
There, the oscillator presents two well defined dynamical regimes, damped and overdamped motions, which typically can be reached by moving a single parameter \cite{calvo2006,pastawski2007}.
In this case, there is an analytical discontinuity in the plot of some dynamical observables, like the oscillation frequency, versus the control parameter.
Due to its similarity with thermodynamical phase transitions, the phenomenon is known as dynamical phase transitions~\cite{calvo2006,pastawski2007}.

In quantum mechanics, the same kind of phenomenon appears. There, the movement of a single parameter of the Hamiltonian may produce abrupt changes (non-analytical) in the decay dynamics of quantum particles or in their spectra.\cite{alvarez2006,dente2008,rotter2010,garmon2012,rotter2015,ruderman2016,garmon2019}

For time-independent Hamiltonians, QDPTs are usually analyzed through the poles of the retarded (or advanced) Green function $G^{r}\left(\varepsilon\right)$ (or $G^{a}\left(\varepsilon\right)$),
\cite{dente2008}
\begin{equation}
\boldsymbol{G}^{r/a}\left(\varepsilon\right)=
\underset{\eta\rightarrow 0^{+}} 
{\lim}\left(\left(\varepsilon\pm i\eta\right) \boldsymbol{I}-\boldsymbol{H}\right)^{-1},
\label{eq:G^{r/a}}
\end{equation}
This is natural as all spectral and dynamical properties of the system can be obtained from the elements of $\boldsymbol{G}^{r}$.
For instance, the local density of states (LDOS) at a site $n$ of a tight-binging Hamiltonian is given by:
\begin{equation}
\begin{array}{ccc}
N_{n}\left(\varepsilon\right) & = & \underset{\eta\rightarrow0^{+}}{\lim}-\frac{1}{\pi}\mathrm{Im}\left\{ G_{nn}^{r}\left(\varepsilon+i\eta\right)\right\}
\end{array}.
\label{eq:LDOS}
\end{equation}
On the other hand, the survival probability of a particle in the $n$-th site,  
$P_{n}\left(t\right)=\left|\left\langle \psi_{n}\left(t\right)\right|\left.
\psi_{n}\left(0\right)\right\rangle \right|^{2}$,  can be expressed as \cite{dente2008}
\begin{equation}
\begin{array}{ccc}
P_{n}\left(t\right) & = & \left|\Theta\left(t\right)\stackrel[-\infty]{\infty}{\varint}
d\varepsilon N_{n}\left(\varepsilon\right)e^{-i\frac{\varepsilon t}{\hbar}}\right|^{2}
\end{array},
\label{eq:P(t)}
\end{equation}
where $\Theta(t)$ is the Heaviside step function.

\subsubsection {Tight-binding Model.\label{sec:TB}}

In this paper, we will consider the tight-binding system depicted in Fig. \ref{fig:1}.
It consists of a single energy level with energy $E_{d}$ coupled, with hopping constants  $V_{L}$ and $V_{R}$, to two semi-infinite tight-binding chains, with site energy $E_{0}$ and hopping $V_{0}$. 
This Hamiltonian represents a minimum model for a quantum dot coupled to two reservoirs and is widely used in the context of quantum transport.
The total Hamiltonian reads
\begin{eqnarray}
\hat{H} & = & \stackrel[n=-\infty]{\infty}{\sum}E^{(n)}\left|\mathit{n}\right\rangle \left\langle n\right|
\notag \\
& &
-V^{(n)}\left(\left|\mathit{n}\right\rangle \left\langle n+1\right|+\left|\mathit{n+1}\right\rangle \left\langle n\right|\right)
\label{eq:H_TB}, 
\end{eqnarray}
where $E^{(0)}=E_{d}$ and $E^{(n)}=E_{0}$  for $n\neq0$, and the hopping constants $V^{(-1)}=V_{L}$, $V^{(0)}=V_{R}$, and $V^{(n)}=V_{0}$ for $n\neq\left\{ -1,0\right\}$.

For convenience, we will define $\hat{H}_{S}$ as the Hamiltonian of the system which includes the first sites of each semi-infinite chain representing the leads.
Then, the three by three matrix $\boldsymbol{H}_{S}$ contains the site energies $E^{(-1)}=E_0$, $E^{(0)}=E_d$, and $E^{(1)}=E_0$, as well the couplings $V_L$ and $V_R$. 

The matrix representation of the eigenvalue equation
$\hat{H}\left|\psi\right\rangle =\varepsilon\left|\psi\right\rangle $, which in principle is of infinite dimension, can be reduced to an effective system of finite dimension by using the decimation technique~\cite{pastawski2001,cattena2014}.
There, the sites connected to leads are corrected by a self-energy associated to the   $\boldsymbol{G}^r(\varepsilon)$
\begin{equation}
 \Sigma_{0}^{r}\left(\varepsilon\right)=\Delta\left(\varepsilon\right)-i\Gamma\left(\varepsilon\right).
\end{equation}
The real ($\Delta$) and imaginary ($\Gamma$) parts of $\Sigma$ are
\begin{equation}
\begin{array}{ccc}
\Delta\left(\varepsilon\right) & = &
\begin{cases}
\frac{\varepsilon-E_{0}}{2}-\sqrt{\left(\frac{\varepsilon-E_{0}}{2}\right)^{2}-V_{0}^{2}}  
& (\mathrm{case \ A})\\
\frac{\varepsilon-E_{0}}{2}
& (\mathrm{case \ B}) \\
\frac{\varepsilon-E_{0}}{2}+\sqrt{\left(\frac{\varepsilon-E_{0}}{2}\right)^{2}-V_{0}^{2}}
& (\mathrm{case \ C})
\end{cases}
\end{array} ,
\end{equation}
\begin{equation}
\begin{array}{ccc}
\Gamma\left(\varepsilon\right) & = &
\begin{cases}
0 
& (\mathrm{case \ A}) \\
\sqrt{V_{0}^{2}-\left(\frac{\varepsilon-E_{0}}{2}\right)^{2}}
& (\mathrm{case \ B}) \\
0 
& (\mathrm{case \ C})
\end{cases}
\end{array},\label{eq:G^r_TB}
\end{equation}
where $\varepsilon-E_{0}\geq 2V_{0}$ for case A, $\varepsilon -E_{0} \in [-2V_{0},2V_{0}]$ for case B, and $\varepsilon - E_{0} \leq -2V_{0}$ for case C. The region where $\Gamma \neq 0$ (case B), the band, corresponds to the energy regions where excitations propagate freely inside the leads.
Using the above self-energy we can define the effective Hamiltonian $\hat{H}^r_{\rm eff} \left(\varepsilon\right) = \hat{H}_{S}+\hat{\Sigma}^r\left(\varepsilon\right)$, where $\hat{\Sigma}^r\left(\varepsilon\right)$ is given by
$
\hat{\Sigma}^r\left(\varepsilon\right)=
\Sigma_{0}^{r}
\left(\varepsilon\right)
\left(
\left|-1\right\rangle
\left\langle -1\right|
+\left|1\right\rangle 
\left\langle 1\right|
\right)
$. 
In matrix form $\hat{H}^r_{\rm eff}$ is 
\begin{equation}
\begin{array}{ccc}
\boldsymbol{H}^r_{\rm eff}(\varepsilon) & = &
\begin{pmatrix}E_{0}+\Sigma_{0}^{r}\left(\varepsilon\right) & -V_{L} & 0\\
-V_{L} & E_{d}-i\Gamma_\eta & -V_{R}\\
0 & -V_{R} & E_{0}+\Sigma_{0}^{r}\left(\varepsilon\right)
\end{pmatrix}\end{array}.
\label{eq:H_eff}
\end{equation}
By means of Eq. \ref{eq:G^{r/a}}, $\boldsymbol{H}^r_{\rm eff}(\varepsilon)$ can be used to calculate $\boldsymbol{G}^r$, while $\boldsymbol{G}^a=\left [\boldsymbol{G}^r\right ]^\dagger$.
The resulting eigenvalue equation $\left(\varepsilon\boldsymbol{I}-\boldsymbol{H}^r_{\rm eff}\left(\varepsilon\right)\right)\left|\psi\right\rangle =0$ is now finite and exact.
The price to be paid is that now the coefficients of $\boldsymbol{H}^r_{\rm eff}$ are energy-dependent.
Notice that, because $\Sigma^r \in \mathbb {C}$, the effective Hamiltonian becomes non-Hermitian,
and thus, its eigenvalues, or equivalently, the poles of the associated Green's function $\boldsymbol{G}^r$ are not necessarily real.
Importantly, these poles can change abruptly with the parameters of the Hamiltonian and, in consequence, also the LDOS [see panels from (c) to (e) in Fig. \ref{fig:2} ] and the dynamics of the system [see panels from (f) to (h) in Fig. \ref{fig:2} ].

Finally, one last detail. 
Due to technical reasons, we perturbatively couple a third lead to the system, see $\Gamma_\eta$ in Eq. \ref{eq:H_eff}.
This reservoir is modeled by a self-energy in the wideband approximation $\Sigma_{\eta}=-i\Gamma_{\eta}$ with $\Gamma_{\eta}$ positive. 
The purpose of this extra reservoir is to add a small broadening to the LDOS and to guarantee an occupation to the system.
Both effects are particularly important when the eigenenergies of the effective Hamiltonian lay outside the band or when $V_L=V_R=0$.
Despite this, in all calculations, we take the limit $\Gamma_{\eta}\rightarrow 0$, and thus its effects on the poles structure is negligible.

\subsubsection{Poles and Dynamical Phase Transitions.\label{sec:poles}}

The poles of the system correspond to the singular points of the Green's function.
The specific tight-binding model described in the previous section has two poles $\left\{ \varepsilon_{p,+},\varepsilon_{p,-}\right\}$ given by the solutions of the secular equation
$\det\left(\varepsilon\boldsymbol{I}-\boldsymbol{H}_{\rm eff}\right)=0$.
Only three variables are really necessary to describe such poles, $E_{d}$, $V_{L}$, and $V_{R}$.
The effect of $E_0$ is only to shift the poles' position while $V_0$ gives the energy scale. For that reason, we define the dimensionless parameters
\begin{eqnarray}
\epsilon_{p,\pm} =  \frac{\varepsilon_{p,\pm}-E_{0}}{V_{0}} , \quad \epsilon_d = \frac{E_{d}-E_{0}}{V_{0}}
\notag \\
\mathrm{and} \quad v^{2} = v_L^2+ v_R^2. 
\label{eq:dimless_param}
\end{eqnarray}
where $v_L=\frac{V_{L}}{V_{0}}$ and $v_R=\frac{V_{R}}{V_{0}}$.
The real and imaginary part of the dimensionless poles are, respectively,\cite{bustos2010,garmon2012}
\begin{eqnarray}
\mathrm{Re}\left(\epsilon_{p,\pm}\right) & = &
\begin{cases}
\epsilon_{d}+\frac{v^{2}}{\left(1-v^{2}\right)}\left[\frac{\epsilon_{d}}{2}\pm\sqrt{\left(\frac{\epsilon_{d}}{2}\right)^{2}+v^{2}-1}\right] 
\\
\epsilon_{d}+\frac{v^{2}}{\left(1-v^{2}\right)}\frac{\epsilon_{d}}{2}
\end{cases}
\notag \\
\mathrm{Im}\left(\epsilon_{p,\pm}\right) & = &
\begin{cases}
0
\\
\pm\frac{v^{2}}{\left(1-v^{2}\right)}\sqrt{1-\left[ \left(\frac{\epsilon_{d}}{2}\right)^{2}+v^{2} \right] }
\end{cases},
\label{eq:poles}
\end{eqnarray}
for $v^{2}\neq1$. The first case of $\mathrm{Re}\left(\epsilon_{p,\pm}\right)$ and $\mathrm{Im}\left(\epsilon_{p,\pm}\right)$ corresponds to the condition $\left(\frac{\epsilon_{d}}{2}\right)^{2}+v^{2}\geq1$, while the other one corresponds to the condition $\left(\frac{\epsilon_{d}}{2}\right)^{2}+v^{2}\leq1$

\begin{figure}
\begin{centering}
\includegraphics[width=3.2in]{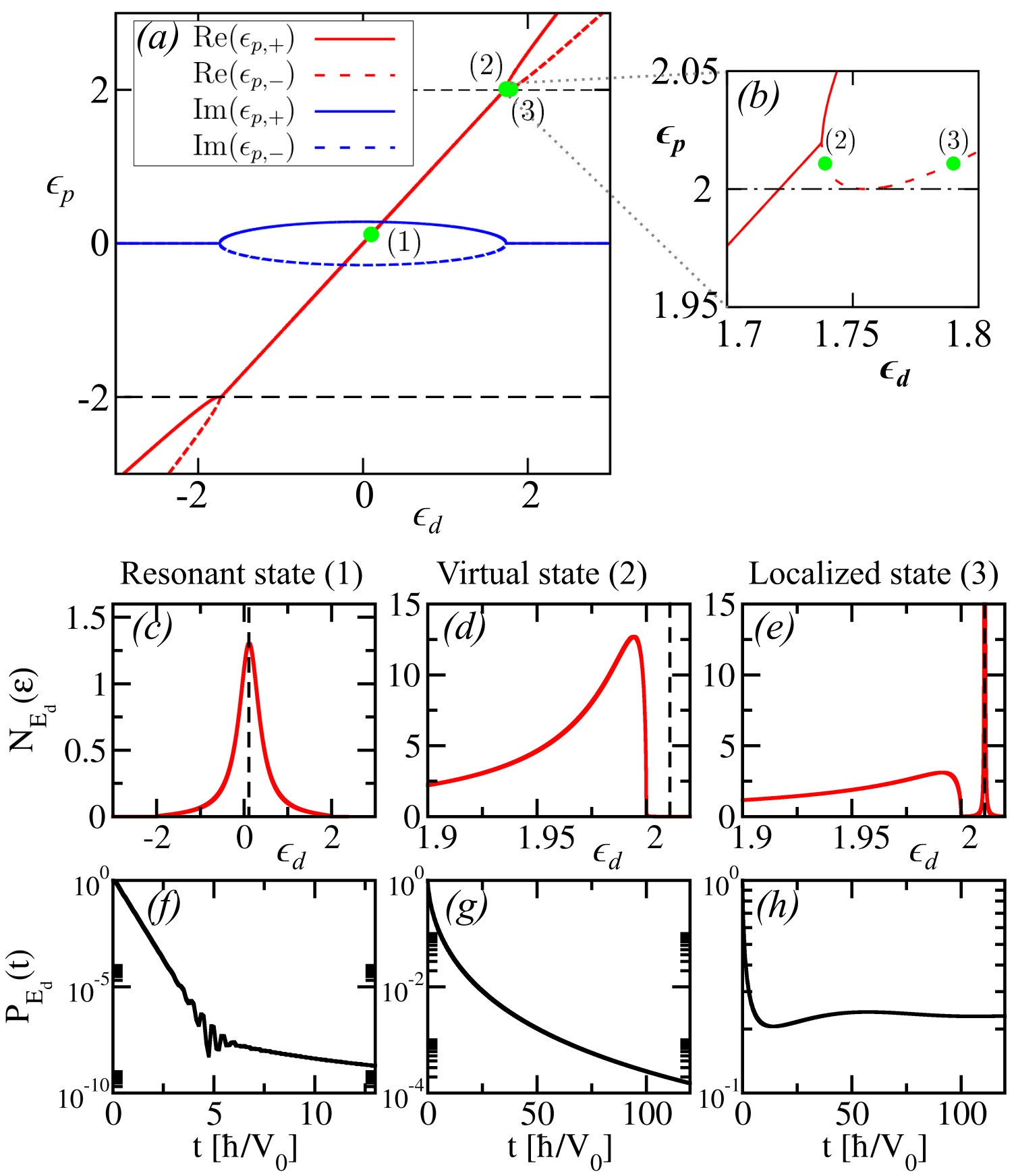}
\par\end{centering}
\caption{\textbf{(a) -} Real [$\mathrm{Re}(\epsilon_{p,\pm})$] and imaginary [$\mathrm{Im}(\epsilon_{p,\pm})$] parts of the poles of the system as function of the dot's energy $\epsilon_d$. 
Black dashed lines show the band edges. The green dots mark three examples of poles. Dot (1) corresponds to a resonant state, (2) to a virtual state, and (3) to a localized state.
\textbf{(b) -} Detail of panel (a).
\textbf{(c), (d), and (e) -} LDOS for cases (1), (2), and (3) respectively.
Black lines in this case correspond to the real part of the poles for each case.
\textbf{(f), (g), and (h) -}
Survival probability for cases (1), (2) y (3) respectively.
In all plots we used $v_{L}=v_{R}=0.35$.
}
\label{fig:2}
\end{figure}

In Fig. \ref{fig:2}-(a) and (b) we show the real and imaginary part of the poles, $\epsilon_{p,\pm}$, as a function of the dimensionless parameter $\epsilon_d$.
In the lower panels of the figure, see panels (c)-(h), we show the typical spectral and dynamical characteristic of three types of poles defining the resonant state, the virtual state, and the localized state.

\textit{Resonant states}. They correspond to poles with a nonzero imaginary part. See the green dot marked as (1) in Fig. \ref{fig:2}-(a). When the system presents such a pole, the LDOS is typically a Lorentzian peak centered at approximately $\mathrm{Re}(\epsilon_p)$ and whose width is given by $\mathrm{Im}(\epsilon_p)$. This description is particularly accurate when $\mathrm{Re}(\epsilon_p)$ is close to the center of the band. However, when $\mathrm{Re}(\epsilon_p)$ approaches one of the band edges the maximum of the LDOS is shifted toward the closest band edge and its shape deviated from a perfect Lorentzian function.
After a short quadratic decay, the dynamics of such systems typically shows an exponential decay.
\cite{fiori2006}

\textit{Localized states}. They correspond to poles with $\mathrm{Im}(\epsilon_p)=0$ and where the $\mathrm{Re}(\epsilon_p)$ lays outside the band ($|\mathrm{Re}(\epsilon_p)|>2$). For instance, see the green dot marked as (3) in Fig. \ref{fig:2}-(b).
When the system presents such a pole, its  LDOS typically shows a very narrow peak outside the band (a Dirac's delta in the limit $\Gamma_\eta \rightarrow 0$). The survival probability may show a small decay at short times but then it remains constant with time.

\textit{Virtual states}.
They also correspond to poles with $\mathrm{Im}(\epsilon_p)=0$ and $|\mathrm{Re}(\epsilon_p)|>2$, see the green dot marked as (2) in Fig. \ref{fig:2}-(b).
However, they are poles of the nonphysical Riemann sheet of $\Sigma^r$, see Refs.~\onlinecite{dente2008} and~\onlinecite{bustos2010}. for more details. Due to this, there is not a peak in the LDOS at $\mathrm{Re}(\epsilon_p)$.
However, the effect of these poles on the LDOS within the band is the same as that of localized states. 
There is an accumulation of states near the closest band edge, which grows as the pole approaches this band edge. See the examples shown in panels (d) and (e).

The three type of poles described above determine two QDPT namely the \emph{resonant-virtual} and the \emph{virtual-localized} transitions.
The \emph{Resonant-virtual} QDPT is given by~\cite{bustos2010,garmon2012}
\begin{equation}
\left(\frac{\epsilon_d}{2}\right)^{2}+v^{2}=1, \label{eq:RV_QDPT}
\end{equation}
while the \emph{Virtual-localized} QDPT is given by~\cite{bustos2010,garmon2012}
\begin{equation}
v^{2}=2\mp \epsilon_d \label{eq:VL_QDPT}.
\end{equation}
A more detailed classification may lead to another QDPT as resonant states can be subdivided into two additional type of poles, see Ref.~\onlinecite{dente2008,bustos2010}. However, for the purpose of this work this subclassification is irrelevant and will be ignored.

One final remark is still in order. In more general scenarios of a quantum dot coupled to $m$ identical semi-infinite chains with coupling $V_{i}$, the corresponding poles are still given by Eq.  \ref{eq:poles}, but the quadratic coupling $v^{2}$ should be replaced by
\begin{equation}
\begin{array}{ccc}
v^{2} & = & \left(\frac{V_{1}}{V_{0}}\right)^{2}+\left(\frac{V_{2}}{V_{0}}\right)^{2}+\cdots+\left(\frac{V_{m}}{V_{0}}\right)^{2}\end{array}.\label{eq:v^2}
\end{equation}

\subsection{Current-induced forces.\label{sec:CIFs}}

When a current of quantum particles is coupled to a mesoscopic mechanical device, the classical dynamics of the mechanical part can be described by an effective Langevin equation ~\cite{bode2011,bustos2019}:
\begin{equation}
\begin{array}{ccc}
M_{\nu}\ddot{X}_{\nu}+\frac{\partial U}{\partial X_{\nu}} & = & \mathcal{F}_{\nu}+\xi_{\nu}\end{array}.\label{eq:Langevin}
\end{equation}
Here, on the left-hand side we have the terms corresponding to the classical degrees of freedom, where the coordinates  $\left\{ X_{\nu} \right\}$ are associated with masses $M_{\nu}$, and $\frac{\partial U}{\partial X_{\nu}}$ accounts for any external forces. 
The terms on the right-hand side of Eq. (\ref{eq:Langevin}) give the CIFs, or the forces arising from the interaction between the mechanical device and the quantum particles (electrons in our case). 
The term $\xi_{\nu}$ describes the quantum fluctuations of the CIFs while $\mathcal{F}_{\nu}$ is the mean value of the force operator
\begin{equation}
\begin{array}{ccc}
\mathcal{F}_{\nu} & = & \left\langle -\frac{\partial\hat{H}}{\partial X_{\nu}}\right\rangle \end{array},\label{eq:<F>}
\end{equation}
where $\hat{H}$ is the electronic Hamiltonian.

\subsubsection{CIFs in the Keldysh Formalism.\label{sec:CIFs_keldysh}}

The mean value of the force operator can be calculated in terms of Green's functions by
\begin{equation}
\begin{array}{ccc}
\mathcal{F}_{\nu} & = & \int\frac{d\varepsilon}{2\pi i}\mathbf{\textrm{Tr}}\left[
\boldsymbol{\Lambda}_{\nu}
\boldsymbol{\mathcal{G}}^{<}\right]
\end{array},\label{eq:F_definition}
\end{equation}
where $\boldsymbol{\Lambda}_{\nu}=-\frac{\partial \boldsymbol{H}}{\partial X_{\nu}}$ and $\boldsymbol{\mathcal{G}}^{<}$ is the lesser Green's function.
This function, which for shortness we omitted its energy and time dependence ($\boldsymbol{\mathcal{G}}^{<}\equiv \boldsymbol{\mathcal{G}}^{<}(\varepsilon,\tau)$), is indeed the Wigner transform of what generalizes the density matrix and is called the particle propagator $\mathcal{G}^{<}(x,t;x',t')= (i/\hbar) \left < \psi^\dagger(x',t') \psi(x,t)\right >$.\cite{haug2008,bode2011}
The closely related Green function $\mathcal{G}^{>}(x,t;x',t')= -(i/\hbar) \left < \psi(x,t) \psi^\dagger(x',t')\right >$ is dubbed the hole propagator, since the order of the creation and annihilation operators are reversed.
Both functions are directly linked to observables and kinetic properties, such as particle densities and currents.
\cite{pastawski1992,haug2008}

In complex scenarios like the one treated here, where the time-dependent Hamiltonians are slowly varying functions of some parameters, the function $\boldsymbol{\mathcal{G}}^{<}$ can be calculated by resorting to an adiabatic expansion in terms of the time-dependent parameters. \cite{haug2008,pastawski1992}
There, $\boldsymbol{\mathcal{G}}^{<}$ is given, up to first order, by, \cite{bode2011}
\begin{eqnarray}
\boldsymbol{\mathcal{G}}^{<} & \simeq & \boldsymbol{G}^{<} -\frac{i \hbar }{2}\underset{\nu}{\sum}\dot{X_{\nu}}\left[\left(\partial_{\varepsilon}\boldsymbol{G}^{<}\right)
 \boldsymbol{\Lambda}_{\nu}
 G^{a}-G^{r}\boldsymbol{\Lambda}_{\nu}\left(\partial_{\varepsilon}\boldsymbol{G}^{<}\right)\right. \notag \\
 &  & \left.+\left(\partial_{\varepsilon}G^{r}\right)\boldsymbol{\Lambda}_{\nu}\boldsymbol{G}^{<}-\boldsymbol{G}^{<}\boldsymbol{\Lambda}_{\nu}\left(\partial_{\varepsilon}G^{a}\right)\right]
\label{eq:G^<adiabatica}
\end{eqnarray}
where $\boldsymbol{G}^{<}$ is the frozen lesser Green's function, given by
\begin{eqnarray}
\boldsymbol{G}^{<} & = & \boldsymbol{G}^{r}\boldsymbol{\Sigma}^{<}\boldsymbol{G}^{a}. \label{eq:G^<}
\end{eqnarray}
Here, $\boldsymbol{G}^{r/a}$ are given by Eq. \ref{eq:G^{r/a}} and $\boldsymbol{\Sigma}^{<}$ is the lesser self-energy, $\boldsymbol{\Sigma}^{<}=2i\underset{\alpha}{\sum}f_{\alpha}\boldsymbol{\Gamma}_{\alpha}$.
The latter contains the information of the reservoirs, where $f_{\alpha}$ is the Fermi-Dirac distribution function of reservoir $\alpha$ with chemical potential $\mu_{\alpha}$ and temperature $T_\alpha$.
The function $\Gamma_{\alpha}$ is the imaginary part of the self-energy of the reservoir $\alpha$, and describes the escape rate from the region of the contact in units of $\hbar$.
Importantly, in deriving Eq. \ref{eq:G^<adiabatica} it was assumed that the self-energies $\boldsymbol{\Sigma}^{<}$ and $\boldsymbol{\Sigma}^{r/a}$ are time-independent. 
However, as we included the first sites of the leads into the effective Hamiltonian, Eq. \ref{eq:H_eff},
a variation of the couplings $V_L$ and $V_R$ does not change the self-energies.

Then, under the adiabatic approximation, $\mathcal{F}_{\nu}$ is
\begin{equation}
\begin{array}{ccc}
\mathcal{F}_{\nu} & \simeq & F_{\nu}-\underset{\nu'}{\sum}\gamma_{\nu\nu'}\dot{X}_{\nu'}\end{array},\label{eq:F_adiabatica}
\end{equation}
where the ``frozen'' force or, from now on, simply the force, is
\begin{equation}
\begin{array}{ccc}
F_{\nu} & = & \int\frac{d\varepsilon}{2\pi i}\mathbf{\textrm{Tr}}\left[\boldsymbol{\Lambda}_{\nu}\boldsymbol{G}^{<}\right]\end{array}.
\label{eq:F_frozen}
\end{equation}
The terms $\gamma_{\nu\nu'}$ are the components of the electronic-friction tensor $\boldsymbol{\gamma}$.
The diagonal elements of $\boldsymbol{\gamma}$ provide the mechanical dissipation while the nondiagonal components are Lorentz-like terms which allow energy transfer between modes.
This tensor can be linearly decomposed into a symmetric and an antisymmetric contribution
$\boldsymbol{\gamma}=\boldsymbol{\gamma}^{s}+\boldsymbol{\gamma}^{a}$. 
The antisymmetric component is finite only for nonzero voltage biases or temperature gradients between the reservoirs. Thus, close to equilibrium conditions only the symmetric contribution of the electronic-friction tensor needs to be taken into account.

\subsubsection{Expansion of the CIFs.\label{sec:CIFs_expansion}}

In this work, we are interested only on the leading orders of the expansion of $\mathcal{F}$ in terms of
the different nonequilibrium sources $\delta \mu$, $\delta T$, and $\dot{\overrightarrow{X}}$,
\begin{eqnarray}
\mathcal{F}_{\nu} & \approx & \left.\mathcal{F}_{\nu}\right|_\mathrm{eq}+\sum_{\nu'}\left.\frac{\partial\mathcal{F}_{\nu}}{\partial\dot{X}_{\nu'}}\right|_{\mathrm{eq}}\dot{X}_{\nu'}
\notag \\
 &  & +\sum_{\alpha}\left.\frac{\partial\mathcal{F}_{\nu}}{\partial\mu_{\alpha}}\right|_{\mathrm{eq}} \delta \mu_{\alpha}
 +\sum_{\alpha}\left.\frac{\partial\mathcal{F}_{\nu}}{ \partial T_{\alpha}}\right|_{\mathrm{eq}} \delta T_{\alpha} .
 \label{eq:F_expans}
\end{eqnarray}
The equilibrium contribution of the force,
\begin{equation}
F_{\nu}^{\mathrm{eq}} = \left . \mathcal{F}_\nu \right |_{\mathrm{eq}},
\end{equation}
is conservative and thus can be written as the gradient of a potential, $\overrightarrow{F}^\mathrm{eq} = -\nabla U^\mathrm{eq}$.
As we show in appendix \ref{app:U^eq}, the equilibrium potential $U^\mathrm{eq}$ is
\begin{equation}
\begin{array}{ccc}
U^\mathrm{eq} & = & -\int\frac{d\varepsilon}{2\pi i}f_{0}\left(\varepsilon\right)\ln\left[\frac{\det\left(\boldsymbol{G}^{a}\right)}{\det\left(\boldsymbol{G}^{r}\right)}\right]
\end{array},\label{eq:U_eq}
\end{equation}
where $f_0$ is the equilibrium Fermi-Dirac distribution function given by $\mu_0$ and $T_0$, the equilibrium chemical potential and temperature respectively.

Comparing Eqs. \ref{eq:F_expans} and \ref{eq:F_adiabatica}, it is obvious that $\left.\frac{\partial\mathcal{F}_{\nu}}{\partial\dot{X}_{\nu'}}\right|_{\mathrm{eq}} = \gamma_{\nu,\nu'}|_\mathrm{eq}$.
Therefore, only the equilibrium contribution of the symmetric component of $\boldsymbol{\gamma}$ is relevant for the present work, where $\gamma_{\nu\nu'}^{s,\mathrm{eq}} \equiv \gamma_{\nu\nu'}^{\mathrm{eq}}$ and
\begin{equation}
\begin{array}{ccc}
\gamma_{\nu\nu'}^{\mathrm{eq}} & = & \int\frac{\hbar d\varepsilon}{4\pi}\mathbf{\textrm{Tr}}\left[\left(\boldsymbol{\Lambda}_{\nu}\boldsymbol{G}^{<}_\mathrm{eq}
\boldsymbol{\Lambda}_{\nu'}
+\boldsymbol{\Lambda}_{\nu'}\boldsymbol{G}^{<}_\mathrm{eq }\boldsymbol{\Lambda}_{\nu}\right)\partial_{\varepsilon}\boldsymbol{G}^{>}_\mathrm{eq}\right]
\end{array}.
\label{eq:gamma^s}
\end{equation}
Here, $\boldsymbol{G}^{\lessgtr}_\mathrm{eq}$ are the equilibrium greater ($>$) and lesser ($<$) Green's functions.
The above equation can be rewritten, see appendix \ref{app:gamma}, in a simpler form,
\begin{equation}
 \gamma_{\nu\nu'}^{\mathrm{eq}} =
 \int\frac{\hbar d\varepsilon}{4\pi}\left(-\partial_{\varepsilon}f_{0}\right)
 \mathbf{\textrm{Tr}} \left[\boldsymbol{\Lambda}_{\nu}\boldsymbol{A}\boldsymbol{\Lambda}_{\nu'}\boldsymbol{A}\right] ,
 \label{eq:gamma_T_finite}
\end{equation}
where $\boldsymbol{A}$ is the spectral function, $\boldsymbol{A}=i\left(\boldsymbol{G}^{r} - \boldsymbol{G}^{a}\right)$.
This equation shows that only states with energies close to the Fermi energy contribute to the equilibrium electronic-friction tensor.
Those states are also responsible for the electric current \cite{pastawski2001} and the nonequilibrium contribution of the CIFs, as we will see afterward.
Therefore, Eq. \ref{eq:gamma_T_finite} highlights a fundamental physical connection between these quantities. 
In the low temperature limit, Eq. \ref{eq:gamma_T_finite} becomes
\begin{eqnarray}
 \underset{k_{B}T_{0}\rightarrow0^{+}}{\lim}\gamma_{\nu\nu'}^{\mathrm{eq}} & = & \frac{\hbar}{4\pi}\mathbf{\textrm{Tr}}
 \left[\boldsymbol{\Lambda}_{\nu}\boldsymbol{A}\boldsymbol{\Lambda}_{\nu'}\boldsymbol{A}\right]_{\varepsilon=\mu_0}. \label{eq:gamma^eq}
\end{eqnarray}

Stochastic forces, in Eq. \ref{eq:Langevin}, can be obtained by using the above equations and the fluctuation-dissipation relation,~\cite{bode2011} given by
\begin{equation}
D_{\nu\nu'}^\mathrm{eq}=2k_{B}T_0 \gamma_{\nu\nu'}^\mathrm{eq},
\end{equation}
where $k_{B}$ is the Boltzmann constant and it is assumed that stochastic forces are described by
$\left\langle \xi_{\nu}\left(t\right)\xi_{\nu'}\left(t'\right)\right\rangle =D_{\nu\nu'}^\mathrm{eq}\delta\left(t-t'\right)$.

At low voltage biases and temperature gradients, the nonequilibrium forces are given by
\begin{equation}
F_{\nu}^{\mathrm{ne}}= \sum_\alpha
\left. \frac{\partial \mathcal{F}_{\nu}}{\partial T_\alpha} \right |_{\mathrm{eq}} \delta T_{\alpha}
+
\left. \frac{\partial \mathcal{F}_{\nu}}{\partial \mu_\alpha} \right |_{\mathrm{eq}} \delta \mu_{\alpha},
\end{equation}
which in the low-temperature limit (see appendix \ref{app:Fne}) gives
\begin{eqnarray}
F_{\nu}^{\mathrm{ne}} & = & 
\sum_\alpha
\frac{\pi}{3} \left (k_{B}T_{0}\right )^{2}
\mathbf{\textrm{Tr}}\left[\boldsymbol{\Lambda}_{\nu}
\left. \frac{\partial 
\left ( \boldsymbol{G}^{r}\Gamma_{\alpha}\boldsymbol{G}^{a} \right )
}{\partial\varepsilon}\right|_{\varepsilon=\mu_{0}}
\right] \frac{\delta T_{\alpha}}{T_{0}} 
\notag \\ & & 
+
\frac{1}{\pi} \mathbf{\textrm{Tr}}
\left[\boldsymbol{\Lambda}_{\nu}
\left (\boldsymbol{G}^{r} \Gamma_{\alpha}\boldsymbol{G}^{a}\right)_{\varepsilon=\mu_{0}}
\right]\delta\mu_{\alpha}
.
\label{eq:F^ne_TVlow}
\end{eqnarray}
Here, $\delta\mu_{\alpha}=\mu_{\alpha}-\mu_0$, $\delta T_{\alpha}=T_{\alpha}-T_0$, and the expression is only valid for $\delta T_{\alpha} \le T_{0}$.

\subsubsection{Work\label{CIFs_work}}

One of the main motivations of this work is to study the general conditions that maximize the performance of different nanoelectromechanical devices and quantum machines.
In this respect, one of the central quantities to evaluate is the work $W$ done by CIFs during cyclic motions.
If one is dealing with systems where only two or three parameters are required to describe their motion, then it is useful to study the curl of the CIFs, defined as
\begin{eqnarray}
\left(\nabla\times \overrightarrow{F}\right)_{\rho} & = & \partial_{\nu}F_{\nu'}-\partial_{\nu'}F_{\nu},
\label{eq:curl}
\end{eqnarray}
where $\rho$, $\nu$, and $\nu'$ are the indexes of the coordinates and $\nu<\nu'$. 
Now, the work can be obtained from
\begin{equation}
\begin{array}{ccc}
W & = & \underset{S}{\iint}\left(\nabla\times \overrightarrow{F}\right) \cdotp d\overrightarrow{S}
\end{array},\label{eq:W_curl}
\end{equation}
where the integration is done on the surface $S$ enclosed by the closed curve $C$ that describes the cyclic motion.

Now inserting Eq. \ref{eq:F^ne_TVlow} into Eq. \ref{eq:curl}, gives
\begin{eqnarray}
\left(\nabla\times \overrightarrow{F}\right)_{\rho} & = &
\sum_{\alpha}
\frac{\delta\mu_{\alpha}}{\pi i}\mathbf{\textrm{Tr}}\left[\left\{ \boldsymbol{\Lambda}_{\nu}\boldsymbol{A}\boldsymbol{\Lambda}_{\nu'}\boldsymbol{G}^{r}\boldsymbol{\Gamma}_{\alpha}\boldsymbol{G}^{a}\right\} _{a}\right]_{\varepsilon=\mu_{0}}
\notag \\ &  & +
\frac{\pi}{3i}\left ( k_{B}T_{0}\right)^2 \frac{\delta T_{\alpha}}{T_0}
\notag \\ & & \times
\frac{\partial}{\partial\varepsilon}\mathbf{\textrm{Tr}}\left[\left\{ \boldsymbol{\Lambda}_{\nu}\boldsymbol{A}\boldsymbol{\Lambda}_{\nu'}\boldsymbol{G}^{r}\boldsymbol{\Gamma}_{\alpha}\boldsymbol{G}^{a}\right\} _{a}\right]_{\varepsilon=\mu_{0}}
\label{eq:curl_TVsmall}
 \end{eqnarray}
where we used $\boldsymbol{G}^{r/a}\boldsymbol{\Lambda}_{\nu}\boldsymbol{G}^{r/a}=\partial_{\nu}\left(\boldsymbol{G}^{r/a}\right)$,
$\partial_{\nu'}\boldsymbol{\Lambda}_{\nu}=\partial_{\nu}\boldsymbol{\Lambda}_{\nu'}$, and $\left\{ \boldsymbol{\Lambda}_{\nu}..\boldsymbol{\Lambda}_{\nu'}..\right\} _{a}=\left(\boldsymbol{\Lambda}_{\nu}..\boldsymbol{\Lambda}_{\nu'}..-\boldsymbol{\Lambda}_{\nu'}..\boldsymbol{\Lambda}_{\nu}..\right)$.
Eq. \ref{eq:curl_TVsmall} gives the work done per unit area in the low temperatures limit and for small bias voltages or temperature gradients.

\subsubsection{CIFs in our model Hamiltonian\label{sec:CIFs_H_TB}}
\begin{figure}[t]
\begin{centering}
\includegraphics[width=3.2in]{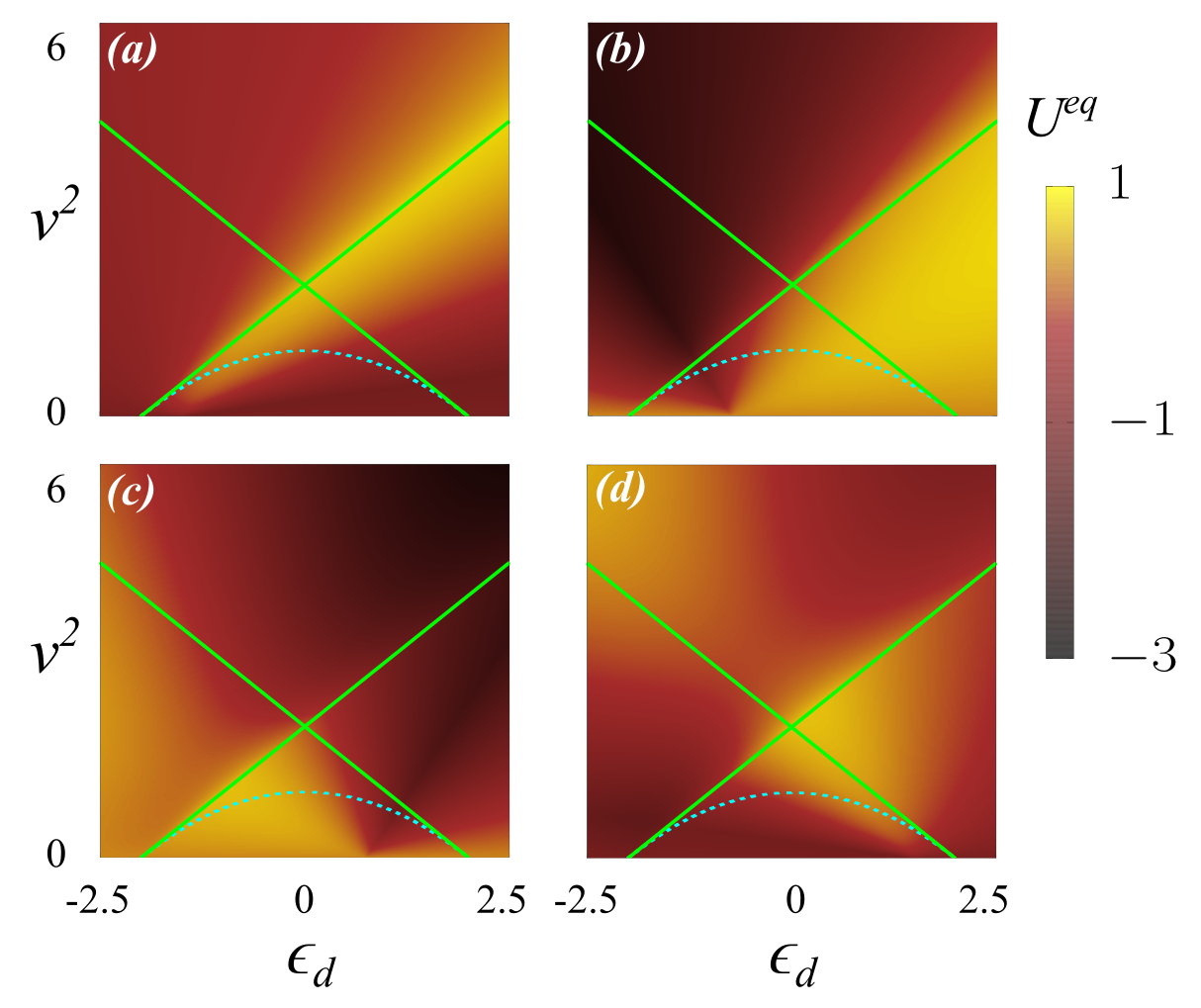}
\par\end{centering}
\caption{Equilibrium potential $U^{\rm eq}$ at $T_{0}=0$ in units of $V_0$, Eq. \ref{eq:U^eq_TB}, for $(a)$ $\epsilon_F=-1.5$, $(b)$ $\epsilon_F=-0.75$, $(c)$ $\epsilon_F=0.75$, and $(d)$ $\epsilon_F=1.5$. Green solid lines indicate the virtual-localized QDPT (Eq. \ref{eq:VL_QDPT}) and cyan dotted lines show the resonant-virtual QDPT (Eq. \ref{eq:RV_QDPT}). }
\label{fig:3}
\end{figure}

Let us return to the tight-binding model described in Sec. \ref{sec:TB}.
For the analysis of the forces, especially for the equilibrium forces, it is crucial to have a definite occupation of the system, even for localized states or when $V_L=V_R=0$.
As discussed in section \ref{sec:poles}, when the system is at the localized state, the poles of $\boldsymbol{G}^r$, or the eigenenergies of $\boldsymbol{H}_\mathrm{eff}$, lay outside the band.
At this condition, the imaginary part of the self-energies, $\Gamma$, is zero. The same happens for $V_L=V_R=0$.
This implies the nonphysical condition that localized states do not contribute to the forces, or even worse, that isolated systems do not have forces at all.
To solve this problem, we added the third reservoir ($\eta$) to the model, assuming its occupation is given by the equilibrium Fermi function $f_0=(f_L+f_R)/2$, and taking the limit $\Gamma_\eta \rightarrow 0$ in all calculations.

Given the Hamiltonian of Eq. \ref{eq:H_eff} and the definition of $\boldsymbol{G}^{r/a}$ (Eq. \ref{eq:G^{r/a}}) and $\boldsymbol{G}^<$ (Eq. \ref{eq:G^<}), we obtain a closed-form expression for the Green functions, (see appendix \ref{app:GFs}).
As can be seen from the expressions of the appendix, the third reservoir does not affect the calculations for energies within the band, provided $\Gamma_\eta$ is small.
However, it allows the calculation of equilibrium $\boldsymbol{G}^{<}$ (and thus equilibrium forces) at all conditions.
Then, all occupied states, independently if they are inside or outside the band, contribute to the equilibrium forces. One important point to emphasize is that the nonequilibrium contribution of $\boldsymbol{G}^{<}$ (and thus $\overrightarrow{F}^{(\mathrm{ne})}$) is always zero for states outside the band (localized states), or for $V_L=V_R=0$. This will have important consequences when studying the maximization of the curl of the forces.

According to Eq. \ref{eq:F_definition}, the forces depend on how the elements of the electronic Hamiltonian are affected by the movement of the mechanical degrees of freedom, but this is system-dependent. Therefore, to gain generality, we take as ``mechanical'' variables the same parameters of the Hamiltonian ($E_{d},V_{L},V_{R}$), which leads to dimensionless forces $F_{\nu}$, see appendix \ref{app:mech_var}.
One can readily calculate the forces in terms of ``physical'' variables $q_{i}$ by
\begin{eqnarray}
F_i & = & \sum_\nu F_\nu \frac{\partial X_{\nu}}{\partial q_{i}} ,
\end{eqnarray}
where $X_{\nu}=\left\{ E_{d}, V_{L}, V_{R}\right\}$. 
Notice that while forces may depend on the chosen variables, the work in Eq. \ref{eq:W_curl} is independent of them, as long as the curve $C$ remains the same, see appendix \ref{app:mech_var}.
Therefore, using our atypical choice of ``mechanical'' variables in Eq. \ref{eq:curl_TVsmall} provides a useful tool to analyze in a general way the performance of nanoelectromechanical devices or quantum machines without having to specify the details of their physical implementation.

\section{Results\label{sec:results}}

\subsection{Equilibrium Potential\label{sec:results_potential}}

Using Eq. \ref{eq:U_eq} and the expression for $G^{r/a}$ obtained in appendix \ref{app:GFs}, we calculate the equilibrium potential 
\begin{eqnarray}
U^\mathrm{eq} & = & -\int\frac{\mathrm{d}\varepsilon}{2\pi i}f_{0}
\ln\left[\frac{\det\left(\boldsymbol{G}^{a}\right )}{\det\left(\boldsymbol{G}^{r}\right )} \right],
\label{eq:U^eq_TB}
\end{eqnarray}
where 
\begin{eqnarray}
\frac{\det\left(\boldsymbol{G}^{a}\right)}{\det\left(\boldsymbol{G}^{r}\right)} & = &
\frac{\left(\epsilon-\widetilde{\Sigma}_{0}^{r}\right)^{2}
\left[\epsilon-\epsilon_d+i\widetilde{\Gamma}_{\eta}-v^{2}\widetilde{\Sigma}_{0}^{r}\right]}
{\left(\epsilon-\widetilde{\Sigma}_{0}^{a}\right)^{2}
\left[\epsilon-\epsilon_d-i \widetilde{\Gamma}_{\eta}-v^{2}\widetilde{\Sigma}_{0}^{a} \right]}.
\label{eq:detGr/detGa_TB}
\end{eqnarray}
Here, we used $\mathrm{det}(\boldsymbol{G}^r)^{-1}=\mathrm{det}(\varepsilon \boldsymbol{I}-\boldsymbol{H}_\mathrm{eff})$ and $\epsilon=(\varepsilon-E_0)/V_0$.
It is worth mentioning that Eq. \ref{eq:U^eq_TB} is also valid for systems coupled to more than two reservoirs where $v^{2}$ is given by Eq. \ref{eq:v^2}.

In Fig. (\ref{fig:3}) we show the potential energy for some representative values of the normalized Fermi energy $\epsilon_{F}= \frac{( \mu_0 -  E_0)}{V_0}$.
The green and cyan lines superimposed to the plot show the different QDPTs.
As can be seen, there is not a clear correlation between them and the equilibrium potential.
However, the figures show a strong dependence of $U^\mathrm{eq}$ with the Fermi energy.
This is important since then, e.g., gate voltages can be used to control the equilibrium position of the system or, even up to some extent, its dynamics.
In this regard, having a simple but general expression for $U^{\mathrm{eq}}$ may result particularly useful.

\subsection{Curl and electronic friction\label{sec:results_curl}}
\begin{figure}
\begin{centering}
\includegraphics[width=3.2in]{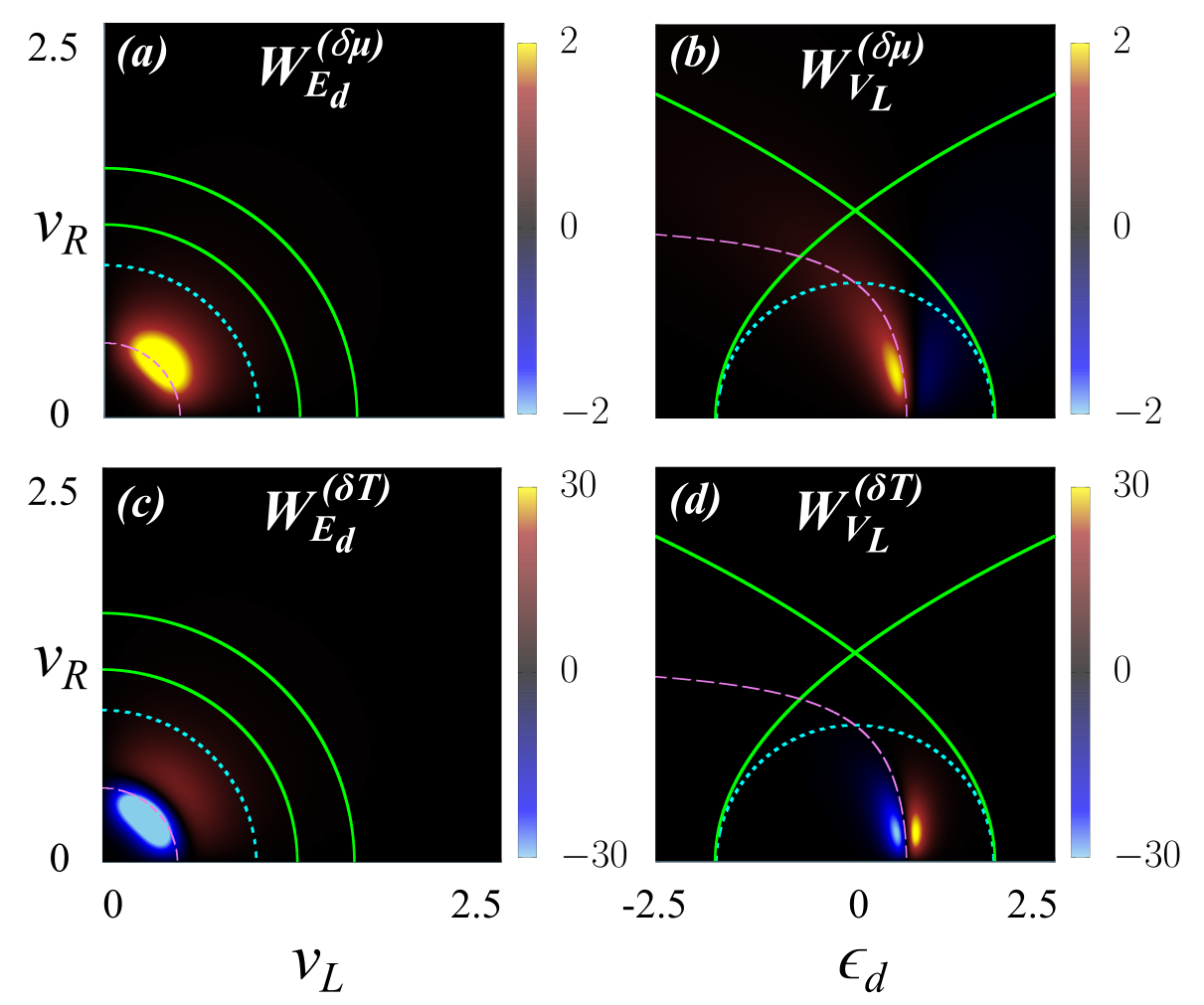}
\par\end{centering}
\caption{Coefficients of the expansion of the curl in terms of $\delta T$ and $\delta \mu$, $W^{(\delta \mu)}$ and $W^{(\delta T)}$ (Eq. \ref{eq:W_definition}), in units of  $1/V_0^2$ and $1/V_0^3$, respectively.
In all cases, $\epsilon_{F}=0.75$, while in $(a)$ and $(c)$, $\epsilon_d=0.5$, and in $(b)$ and $(d)$, $v_L=0.5$
The green solid lines correspond to the virtual-localized QDPTs, the cyan dotted lines show the resonant-virtual QDPTs, and the violet dashed lines indicate the expected condition that should maximize the $W$ coefficients. See the discussion in the main text.}
\label{fig:4}
\end{figure}

The matrices $\boldsymbol{\Lambda}_\nu$, with our choice of mechanical coordinates, see section \ref{sec:CIFs_H_TB}, are 
\begin{eqnarray} 
\boldsymbol{\Lambda}_{V_{L}} & = &
\begin{pmatrix}0 & 1 & 0\\
1 & 0 & 0\\
0 & 0 & 0
\end{pmatrix}
, \qquad
\boldsymbol{\Lambda}_{V_{R}}
=\begin{pmatrix}0 & 0 & 0\\
0 & 0 & 1\\
0 & 1 & 0
\end{pmatrix}
\notag \\
\mathrm{and} & &
\boldsymbol{\Lambda}_{E_{p}} = 
-\begin{pmatrix}0 & 0 & 0\\
0 & 1 & 0\\
0 & 0 & 0
\end{pmatrix}
. \label{eq:Lambda}
\end{eqnarray}
The spectral function can be calculated from $\boldsymbol{A}=2\textrm{Im}\left ( \boldsymbol{G}^{r}\right )$.
Using this in Eq. \ref{eq:curl_TVsmall}, it is possible to obtain a simple expression for the elements of the curl of the forces (see appendixes \ref{app:curl} and \ref{app:curl_TB})
\begin{eqnarray}
\left ( \nabla \right. \times  \overrightarrow{F} \left . \right ) & \simeq &
\overrightarrow{g}\left(\epsilon_{F}\right) \widetilde{T}\left(\epsilon_{F}\right) N_d\left(\epsilon_{F}\right)
\delta \mu  \notag \\
&+&\frac{\pi^{2}}{3}\left.\frac{\partial \left ( \overrightarrow{g}
\widetilde{T} N_d \right) }{\partial\epsilon}\right|_{\epsilon=\epsilon_{F}}\left(k_{B}T_{0}\right)^{2}
\frac{\delta T}{T_{0}}
.\label{eq:curl_TVsmall_TB}
\end{eqnarray}
Here we take $\mu_0 = (\mu_{L}+\mu_{R})/ 2$, $T_{0}= (T_{L}+T_{R})/2$, $\delta \mu=\mu_L-\mu_R$, and $\delta T=T_L-T_R$. The transmittance between reservoirs $L$ and $R$ is $\widetilde{T}\left(\varepsilon\right)$, $N_d\left(\varepsilon\right)$ is the LDOS of the dot, and
\begin{eqnarray}
\overrightarrow{g}= \frac{1}{V_0}
\left(2\frac{ \left (\epsilon-\epsilon_{d}\right ) }{v_L v_R},\frac{1}{v_{R}},\frac{1}{v_{L}}\right).
\label{eq:g} 
\end{eqnarray}
The curl of the force represents the work per unit area where, in our case, the unit area is given in units of energy and therefore the curl has units of one over energy.
\begin{figure}[ht]
\begin{centering}
\includegraphics[width=3.2in]{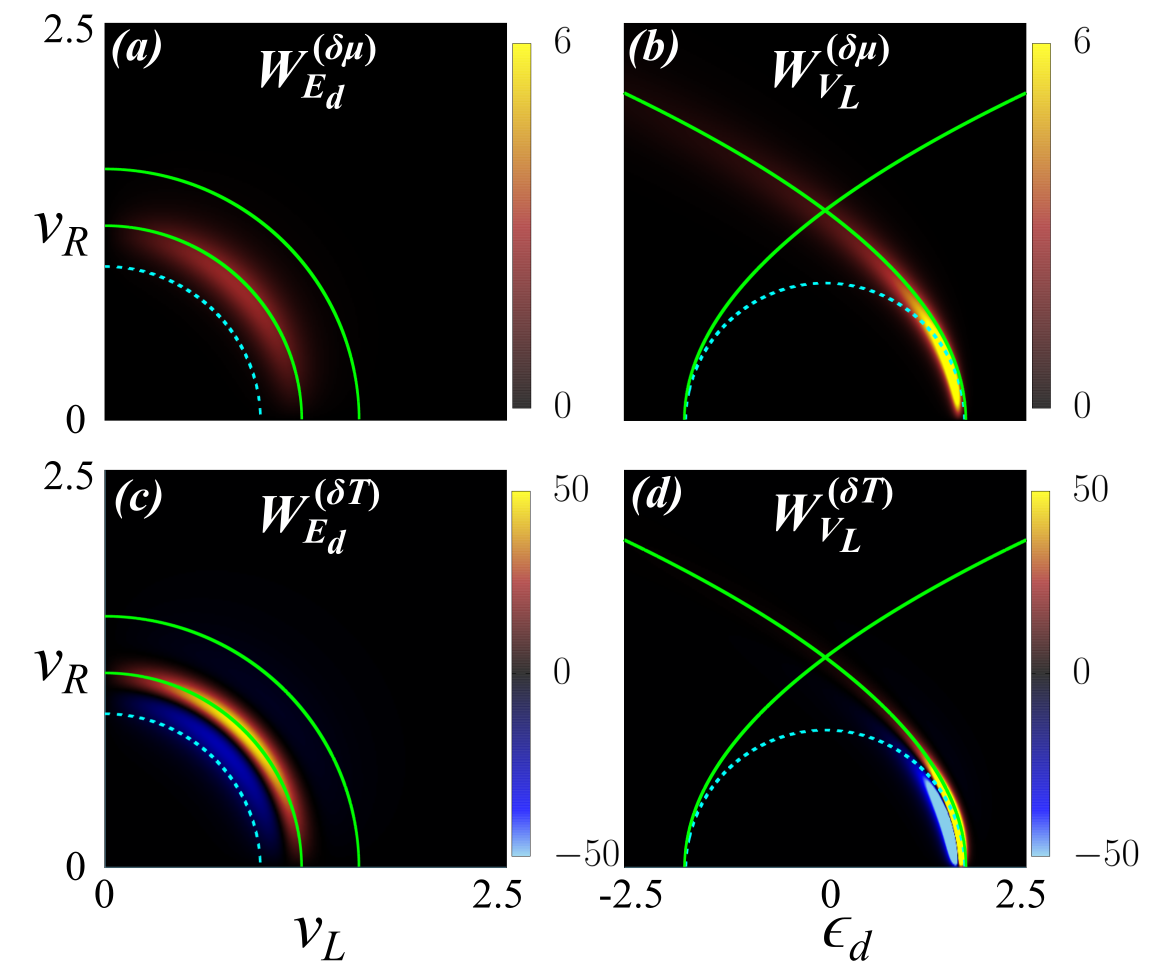}
\par\end{centering}
\caption{Same as Fig. \ref{fig:4} but for $\epsilon_{F}=1.99$.}
\label{fig:5}
\end{figure}

As there are three components of the curl (Eq. \ref{eq:curl_TVsmall_TB}) and two coefficients of the expansion (one for $\delta T$ and one for $\delta \mu$), there are a total of six components of the expansion of the curls. However, due to the symmetry of the problem, the coefficients of the expansion of $(\nabla\times F)_{V_L}$ are equivalent to the coefficients of the expansion of $(\nabla\times F)_{V_R}$ with $v_L$ and $v_R$ interchanged. Therefore, we are going to consider only four coefficients namely $W^{(\delta \mu)}_{E_d}$, $W^{(\delta T)}_{E_d}$, $W^{(\delta \mu)}_{V_L}$, and $W^{(\delta T)}_{V_L}$, where
\begin{eqnarray}
(\nabla\times F)_{V_L} & = &
\left (\delta \mu \right ) W^{(\delta \mu)}_{V_L}
+ 
\left ( k_{B} T_0 \right )^{2} \frac{\delta T}{T_0}
W^{(\delta T)}_{V_L}
\notag \\
(\nabla\times F)_{E_d} & = &
\left (\delta \mu \right ) W^{(\delta \mu)}_{E_d}
+ 
\left ( k_{B} T_0 \right )^{2} \frac{\delta T}{T_0}
W^{(\delta T)}_{E_d} .
\label{eq:W_definition}
\end{eqnarray}
When one multiplies the coefficients $W^{(\delta \mu)}_\rho$ or $W^{(\delta T)}_\rho$ (where $\rho=E_d$, $V_L$, or $V_R$) by $\delta \mu$ or $\left ( k_{B} T_0 \right )^{2} \frac{\delta T}{T_0}$ respectively, the result gives the low temperature limit of the work done per unit area for a cyclic movement of the other two variables.
\begin{figure}
\begin{centering}
\includegraphics[width=3.2in]{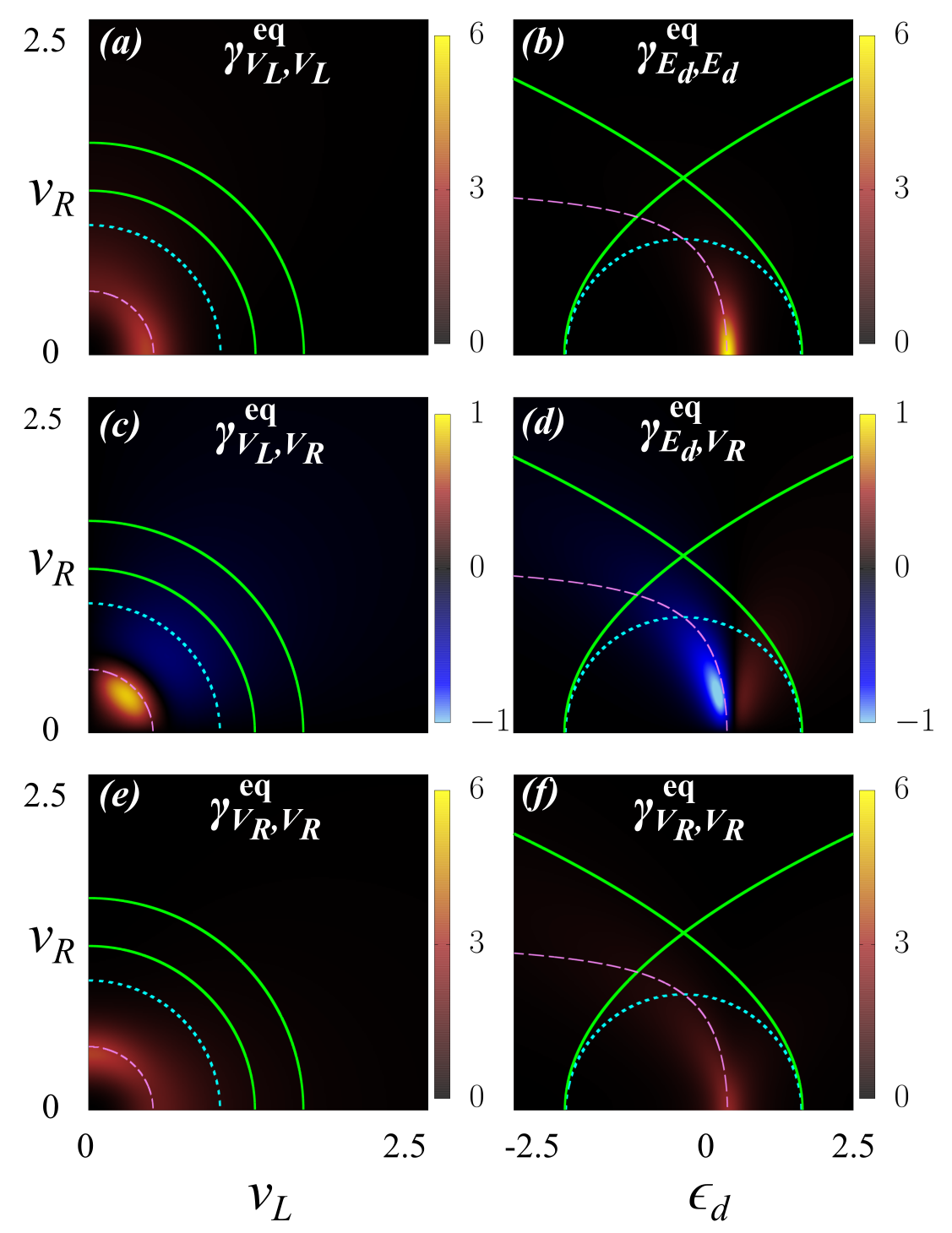}
\end{centering}
\caption{
Elements of the electronic-friction tensor $\boldsymbol{\gamma}$ in units of $\hbar/V_0^2$ for the same condition used in Fig. \ref{fig:4}.
Panels $(a)$, $(c)$, and $(e)$ are the elements of the tensor relevant for the movement of parameters $V_L$ and $V_R$.
Thus, they should be compared with the coefficients $W_{E_d}^{(\delta\mu)}$ and $W_{E_d}^{(\delta T)}$ of Fig \ref{fig:4}.
Similarly panels $(b)$, $(d)$, and $(f)$ should be compared with the coefficients $W_{V_L}^{(\delta\mu)}$ and $W_{V_L}^{(\delta T)}$ of Fig \ref{fig:4}.
In $(a)$, $(c)$, and $(e)$, we used $\epsilon_d=0.5$, while in $(b)$, $(d)$, and $(f)$, $v_L=0.5$.}
\label{fig:6}
\end{figure}

Having a simple expression for the coefficients $W^{(\delta \mu)}_\rho$ or $W^{(\delta T)}_\rho$ may be important when studying different forms of nanoelectromechanical devices, adiabatic quantum motors, or quantum heat engines.
Let us recall that due to Onsager's reciprocal relations, the coefficients of the expansion of the work in terms of $\delta \mu$ and $\delta T$ gives, respectively, the charge pumped by adiabatic quantum pumps and the heat pumped by adiabatic quantum heat pumps~\cite{bustos2013,ludovico2016}.
Therefore, the expressions found for the coefficients and the conclusions regarding their general behavior are also valid for such systems.

In Fig. \ref{fig:4} we show some examples of the behavior of the coefficients of the expansion of the curl. In general, their behavior is highly dependent on the Fermi energy, just as with the case of $U^{\mathrm{eq}}$.
This feature may be useful, e.g., to control the dynamics of a system by using a gate voltage as a knob. Indeed, it is possible to change, in a given region of the parameter's space, the sign of the coefficients, and thus the preferred direction of motion of the system by changing $\epsilon_F$.

The green solid lines and the cyan dashed lines indicate the QDPTs in Fig. \ref{fig:4}.
Although not obvious at first glance, we find that the behavior of the $W$ coefficients is related to the type of poles of the Green's function describing the effective Hamiltonian, Eq. \ref{eq:curl_TVsmall_TB}.
This is so as the shape of the LDOS is determined by the type of pole, which, in turn, depends on the parameters of the Hamiltonian.

The $W_{V_L}$ coefficients depend on the square of the LDOS (or its derivative). Note that, in our system the transmittance $\widetilde{T}\left(\varepsilon\right)$ is proportional to the LDOS within a wide band approximation.\cite{haug2008}
Therefore it is expected a coincidence between the regions with the maximum value of the coefficients and the regions with a maximum value of the LDOS, see Eq. \ref{eq:curl_TVsmall_TB}.
According to the discussions of Sec. \ref{sec:DPT}, for resonant states (see the region enclosed by the cyan dashed line in Fig. \ref{fig:4}), the maximum of the LDOS approximately coincides (for poles not too close to the band edges) with the real part of the pole, see Fig. \ref{fig:2}-(c).
In such a case it is expected a maximum of the $W_{V_L}$ coefficients for $\epsilon_F=\mathrm{Re}(\epsilon_p)$, or
\begin{equation}
 v_R^2= \frac{1}{1+\frac{1}{2 \left ( \frac{\epsilon_F}{E_d}- 1 \right)} }- v_L^2, \label{eq:Max_W_VL}
\end{equation}
where we used Eq. \ref{eq:poles}.
The above equation is plotted in Figs. \ref{fig:4}-(b) and (d) as violet dashed lines. As can be seen, the regions with a maximum value of the $W_{V_L}$ coefficients are close to the line given by Eq. \ref{eq:Max_W_VL}.

For the $W_{E_d}$ coefficients the same analysis can be done except that they depend on the square of the LDOS multiplied by the term $(\epsilon-\epsilon_d$), which shifts the maximum of the function from $\mathrm{Re}(\epsilon_p)$. In this case, the maximum value of the coefficients should be close to the regions where the following function is maximum
\begin{equation}
 \left( \epsilon_F-\epsilon_{d}\right) \left ( \frac{ \mathrm{Im}\left(\epsilon_{p}\right)}{\left[\epsilon_F-\mathrm{Re}\left(\epsilon_{p}\right)\right]^{2}+\mathrm{Im}\left(\epsilon_{p}\right)^{2}}
 \right )^2.
\label{eq:Max_W_Ed}
 \end{equation}
Here, we used the fact that in our system the transmittance is proportional to the LDOS. We assumed the pole is close to the center of the band, and thus $\Gamma$ can be taken in a wide band approximation and the LDOS has a Lorentzian shape centered around $\mathrm{Re}(\epsilon_p)$ and with a width given by $\mathrm{Im}(\epsilon_p)$. And we applied all this to Eq. \ref{eq:curl_TVsmall_TB}. Note that $\epsilon_p$ is a function of $\epsilon_d$, $v_L$, and $v_R$, see Eq. \ref{eq:poles}.
The above condition, numerically found, is plotted in Figs. \ref{fig:4}-(a) and (c) as violet dashed lines.
As can be seen, the regions with a maximum value of the $W_{E_d}$ coefficients are close to this line.
\begin{figure}
\begin{centering}
\includegraphics[width=3.2in]{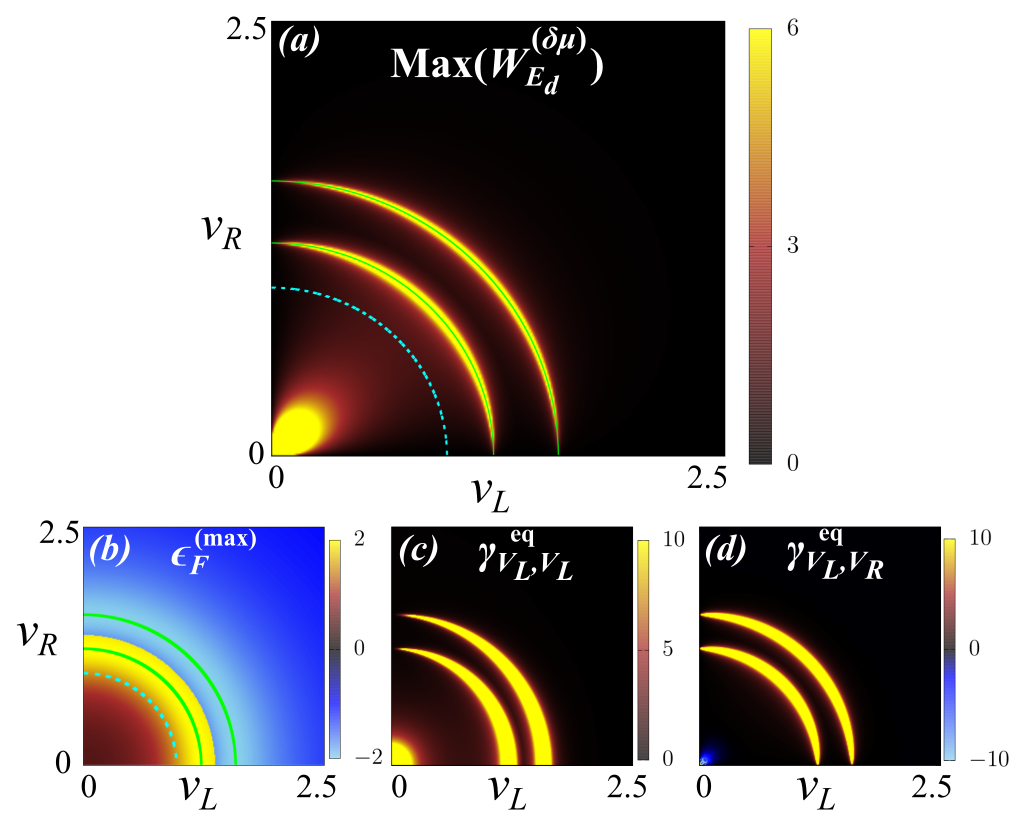}
\par\end{centering}
\caption{
\textbf{(a) -} Maximum value of the coefficient $W_{E_d}^{(\delta\mu)}$ obtained by a variation of the Fermi energy $\epsilon_F$ at every point of the parameter space. Green solid lines show the virtual localized QDPTs and the cyan dotted line indicates the resonant-virtual QDPT.
\textbf{(b) -} Value of $\epsilon_F$ that maximized $W_{E_d}^{(\delta\mu)}$ ($\epsilon_F^{\mathrm{(Max)}}$). 
\textbf{(c) and (d) -} Example of some of the elements of the electronic-friction tensor calculated using $\epsilon_F^{\mathrm{(Max)}}$. 
}
\label{fig:7}
\end{figure}

All the above discussions are valid as long as $|\epsilon_F|\ll 2$. Let us recall that only when $\epsilon_p$ is far from the band edges, the LDOS can be described by a Lorentzian function centered around $\mathrm{Re}(\epsilon_p)$.
When $\mathrm{Re}(\epsilon_p)$ approaches one of the band edges ($|\mathrm{Re}(\epsilon_p)|\approx 2$), the LDOS is largely distorted exhibiting a peak with a maximum shifted towards the closest band edge.
If we move the pole even further, it will correspond to a virtual state (see Fig. \ref{fig:2}-(a)).
For virtual states ($|\mathrm{Re}(\epsilon_p)| \geq 2$ and $\mathrm{Im}(\epsilon_p)=0$), the LDOS shows a maximum almost at the band edges $\epsilon \approx \pm 2$, see Fig. \ref{fig:2}-(d).
The height of this maximum grows until the virtual-localized QDPT is reached (where the maximum is exactly at $\epsilon=\pm 2$) and then it decreases.
Therefore, in all these cases the maximum of the LDOS is expected to approximately coincide with one of the band edges.
This is the reason why for $|\epsilon_F|\approx 2$ the maximum value of the curl coincide approximately with the virtual-localized QDPT, see Fig. \ref{fig:5}.
For localized states, obviously there is also a maximum outside the band, see Fig. \ref{fig:2}-(e).
However, as discussed in section Sec. \ref{sec:CIFs_H_TB}, states with energies outside the band do not contribute to nonequilibrium forces. Note also that the transmittance $\widetilde{T}\left(\varepsilon\right)$ is zero outside the band and thus $(\nabla \times \overrightarrow{F})=0$ according to Eq. \ref{eq:curl_TVsmall_TB}.

In summary, as a rule of thumb when $|\epsilon_F|\ll 2$, the regions of the parameter space with maximum values of the $W_{V_L}$ coefficients are given by Eq. \ref{eq:Max_W_VL}, while for the $W_{E_d}$ coefficients, these regions are shifted and given by the maximum of Eq. \ref{eq:Max_W_Ed}. On the other hand, when $|\epsilon_F| \approx 2$, the regions of the parameter space with the maximum values of both coefficients ($W_{V_L}$ and $W_{E_d}$) approximately coincide with the virtual-localized QDPT, Eq. \ref{eq:VL_QDPT}.

Here, we also obtained closed-form expressions for the elements of the electronic-friction tensor $\boldsymbol{\gamma}$, see appendix \ref{app:friction_TB}.
Fig. \ref{fig:6} shows some elements of $\boldsymbol{\gamma}$ for the same conditions used in Fig. \ref{fig:4}.
As can be seen, in general, the regions in the parameter space that maximize the $W$ coefficients roughly coincide with the regions that maximize the elements of $\boldsymbol{\gamma}$.
This is reasonable since $\boldsymbol{\gamma}$ and $W$ show a similar, although not exactly the same, dependence with the LDOS and the transmittance, see appendix \ref{app:friction_TB}.
Therefore, one can use much the same arguments to explain why some regions of the parameter space present a maximum.

\begin{figure}[ht]
\begin{centering}
\includegraphics[width=3.2in]{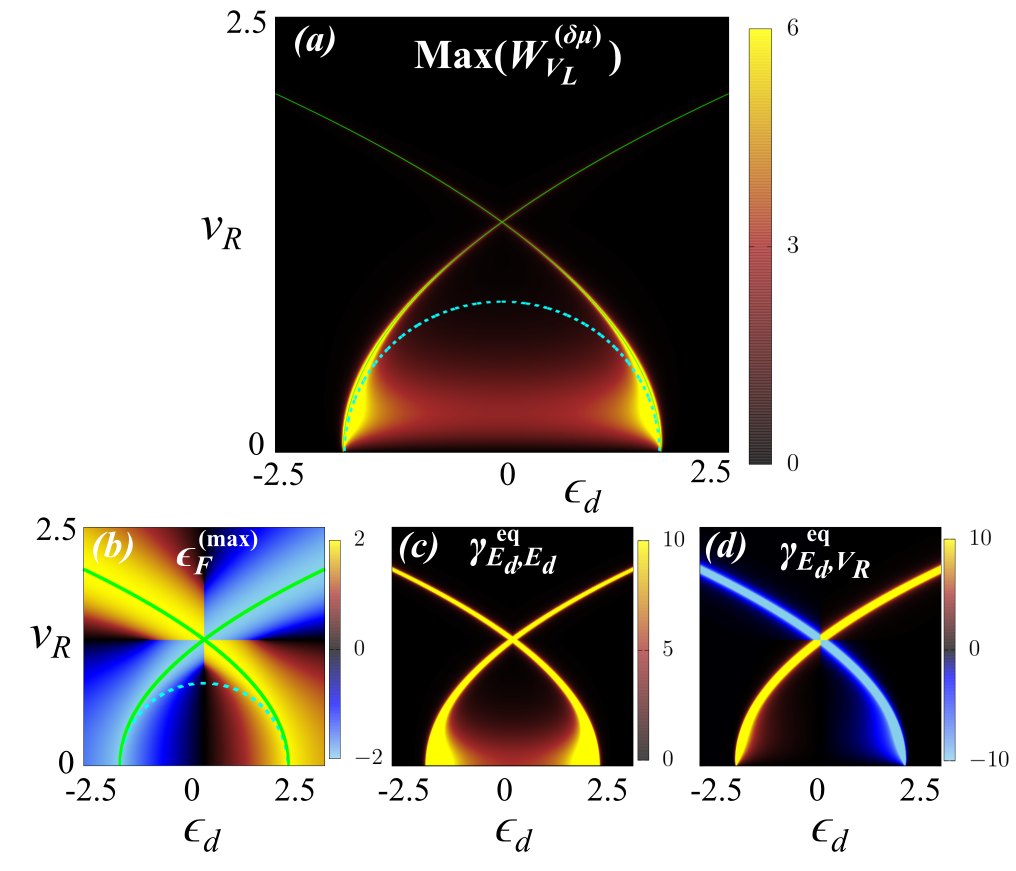}
\par\end{centering}
\caption{
Same as Fig \ref{fig:7} but for $W_{V_L}^{(\delta\mu)}$.
}
\label{fig:8}
\end{figure}

One may wonder, what are the regions in the parameter space that maximizes the $W$ coefficients independently of $\epsilon_F$, i.e., if one could move $\epsilon_F$ at will. This is shown in Figs. \ref{fig:7} and \ref{fig:8}.
We do not show the $W^{(\delta T)}$ coefficients as their behavior is approximately the same as that of the $W^{(\delta \mu)}$ coefficients. 
As can be seen, the regions that maximize the $W$ coefficients coincide in general with the virtual-localized QDPT. Note however that in Fig. \ref{fig:7} the region $v_L \approx v_R \approx 0$ presents values of $W_{E_d}^{(\delta\mu)}$ similar to those of the virtual-localized QDPT.
Regarding the elements of the electronic-friction tensor, they roughly follow the behavior of the $W$ coefficients.

\section{Conclusions\label{sec:conclusions}}

When designing or studying a nanoelectromechanical device or a quantum machine, it is crucial to understand the effect of the system's parameters on their performance. 
Here, we carry out a thorough study of the CIF and the electronic-friction tensor within one of the most common Hamiltonian model in quantum transport.~\cite{haug2008,pastawski2001,zimbovskaya2013}
It is important to point out that any single-particle Hamiltonian, where there is only one relevant state, can be reduced to the model used here by means of a decimation procedure, see, e.g. Ref. \onlinecite{pastawski2001}.

We derived analytic  formulas for the equilibrium potential (Eqs. \ref{eq:U_eq} and \ref{eq:detGr/detGa_TB}), the low-temperature limit of the electronic-friction tensor (Eq. \ref{eq:gamma^eq}), and the low-temperature limit of the curl of the CIFs up to leading order in $\delta T$ and $\delta \mu$ (Eqs. \ref{eq:curl_TVsmall_TB} and \ref{eq:curl_TVsmall}).
We showed the connection between QDPTs, the curl of the CIFs, and the electronic-friction tensor.
Using this as a guide, we found some rule of thumbs that can be useful to quickly identify the regions of the parameters space that should be enclosed to maximize the work done by the device.
Moreover, because of the strategy we used to carry out the analysis, our results are independent of the details of how the mechanical and electronic degrees of freedom are coupled. 

Again, we want to emphasize that, due to Onsager's reciprocal relations, our analysis of the coefficients of the expansion of the curl ($W^{(\delta \mu)}_\rho$ and $W^{(\delta T)}_\rho$) and the formulas shown in Eqs. \ref{eq:curl_TVsmall} and \ref{eq:curl_TVsmall_TB}, are also valid for adiabatic quantum pumps and quantum heat pumps~\cite{bustos2013,ludovico2016}.

The present work may also help to connect the configuration of conducting devices with current-induced structural-failures.
Basically, our simple Hamiltonian also represents a minimal model of an impurity in a one-dimensional conductor.
Therefore, according to our results, some particular values of the parameters of the system should dramatically increase nonconservative forces and the curl of the force, which should ultimately lead to a mechanical failure.
That reasoning, of course, depends on the vibrational modes involved and the dynamical feedback between them and the electronic currents.\cite{lu2010}
However, it is an interesting direction for further research.

Other appealing directions of extending this work include analyzing other common Hamiltonian models, such as double quantum dots, and studying systems within the Coulomb blockade regime of quantum transport.

\section{Acknowledgements} We acknowledge financial support by Consejo Nacional de Investigaciones Cient\'ificas y T\'ecnicas (CONICET); Secretar\'ia de Ciencia y Tecnolog\'ia de la Universidad Nacional de C\'ordoba (SECYT-UNC); and Agencia Nacional de Promoción Científica y Tecnológica (ANPCyT, PICT-2018-03587).

\appendix

\section{Equilibrium potential.\label{app:U^eq}}

By writing the Fermi functions as
$f_{L/R} = f_0 + \Delta f_{L/R}$, one can split the zero order term of the adiabatic expansion of the CIFs $F_{\nu}$ into the equilibrium ($F_{\nu}^{\mathrm{eq}}$) and nonequilibrium ($F_{\nu}^{\mathrm{ne}}$) contributions
\begin{eqnarray*}
F_{\nu}  & = & 
 \int\frac{\mathrm{d}\varepsilon}{\pi}f_{0}\underset{\alpha}{\sum}\mathbf{\textrm{Tr}}\left[\boldsymbol{\Lambda}_{\nu}\boldsymbol{G}^{r}\boldsymbol{\Gamma}_{\alpha}\boldsymbol{G}^{a}\right] \notag \\
 &  & +\int\frac{\mathrm{d}\varepsilon}{\pi}\underset{\alpha}{\sum}\Delta f_{\alpha}\mathbf{\textrm{Tr}}\left[\boldsymbol{\Lambda}_{\nu}\boldsymbol{G}^{r}\boldsymbol{\Gamma}_{\alpha}\boldsymbol{G}^{a}\right]  \notag \\
 & = & F_{\nu}^{\mathrm{eq}}+F_{\nu}^{\mathrm{ne}} .
\end{eqnarray*}
Given their definition, the following relation holds for the Green functions
\begin{eqnarray*}
\left(\boldsymbol{G}^{r}\right)^{-1}-\left(\boldsymbol{G}^{a}\right)^{-1}=2i\underset{\alpha}{\sum}\boldsymbol{\Gamma}_{\alpha}.
\end{eqnarray*}
Therefore, equilibrium force can be written as 
\begin{eqnarray*}
 F_{\nu}^{\mathrm{eq}} & = & \int\frac{\mathrm{d}\varepsilon}{2\pi i}f_{0}
 \mathbf{\textrm{Tr}} \left[ \boldsymbol{\Lambda}_{\nu}    \left ( \boldsymbol{G}^{a}-\boldsymbol{G}^{a} \right )             \right]
\end{eqnarray*}
Now, let us take a matrix $\boldsymbol{U}$ which diagonalize $\boldsymbol{G}^{r}$,
\begin{eqnarray*}
\boldsymbol{U}^{-1}\boldsymbol{G}^{r}\boldsymbol{U} & = & \boldsymbol{D}^{r} .
\end{eqnarray*}
Obviously, $\boldsymbol{U}$ also diagonalize $\left(\boldsymbol{G}^{r}\right)^{-1}$. Then, using $\frac{\partial\boldsymbol{H}}{\partial X_{\nu}}=\frac{\partial\left(\boldsymbol{G}^{r}\right)^{-1}}{\partial X_{\nu}}=\left(\boldsymbol{G}^{r}\right)^{-1}\frac{\partial\boldsymbol{G}^{r}}{\partial X_{\nu}}\left(\boldsymbol{G}^{r}\right)^{-1}$, one finds
\begin{eqnarray*}
\mathbf{\textrm{Tr}}\left[\boldsymbol{\Lambda}_{\nu}\boldsymbol{G}^{r}\right]  =\mathbf{\textrm{Tr}}\left[-\frac{\partial\left(\boldsymbol{D}^{r}\right)^{-1}}{\partial X_{\nu}}\boldsymbol{D}^{r}\right] .
\end{eqnarray*}
The elements of $\boldsymbol{D}^{r}$ are the eigenvalues of $\boldsymbol{G}^{r}$, thus
\begin{eqnarray*}
\mathbf{\textrm{Tr}}\left[-\frac{\partial\left(\boldsymbol{D}^{r}\right)^{-1}}{\partial X_{\nu}}\boldsymbol{D}^{r}\right] & = & -\frac{\partial\ln\left(\det\left(\boldsymbol{G}^{r}\right)\right)}{\partial X_{\nu}} .
\end{eqnarray*}
A similar argument can be used for $\boldsymbol{G}^a$. 
The equilibrium force results in
\begin{eqnarray*}
F_{\nu}^{\mathrm{eq}} & = & \frac{\partial}{\partial X_{\nu}}
\left[
\int\frac{\mathrm{d}\varepsilon}{2\pi i}f_{0}\ln\left(\frac{\det\left(\boldsymbol{G}^{a}\right)}{\det\left(\boldsymbol{G}^{r}\right)}\right)
\right]
,
\end{eqnarray*}
where one can identify the equilibrium potential $U^{\mathrm{eq}}$ as
\begin{eqnarray*}
U^{\mathrm{eq}} & =- & \int\frac{\mathrm{d}\varepsilon}{2\pi i}f_{0}\ln\left(\frac{\det\left(\boldsymbol{G}^{a}\right)}{\det\left(\boldsymbol{G}^{r}\right)}\right)
.
\end{eqnarray*}
Note that $\left|\frac{\det\left(\boldsymbol{G}^{a}\right)}{\det\left(\boldsymbol{G}^{r}\right)}\right|=1$, and then one can write
\begin{eqnarray*}
\frac{\det\left(\boldsymbol{G}^{a}\right)}{\det\left(\boldsymbol{G}^{r}\right)} & = & \exp\left(-i\chi\right)
.
\end{eqnarray*}
Taking $z=\det\left(\boldsymbol{G}^{r}\right)$ and considering $\chi$ as the principal value of the logarithm, gives
\begin{eqnarray*}
\chi & = & \arctan\left(\frac{2\textrm{Re\ensuremath{\left(z\right)}}\textrm{Im}\left(z\right)}{\left(\textrm{Re\ensuremath{\left(z\right)}}\right)^{2}-\left(\textrm{Im\ensuremath{\left(z\right)}}\right)^{2}}\right).
\end{eqnarray*}
Using this, the equilibrium potential can be written as
\begin{eqnarray*}
U^{\mathrm{eq}} & = & \int \frac{\mathrm{d}\varepsilon}{2\pi}f_{0} \chi\left ( \varepsilon \right ) .
\end{eqnarray*}
Note that it is not necessary to integrate from $-\infty$ to $\infty$ but only over energies where $\mathrm{Im}(\boldsymbol{G}^r)\neq 0$ (where $\chi (\varepsilon)\neq 0$).
As the function ``$\arctan$'' is bounded and monotonically increasing, the following inequalities hold
\begin{eqnarray*}
\pi & \geq  \chi \geq & -\pi  \notag \\
\frac{1}{2}\int_{*}\mathrm{d}\varepsilon f_{0} & \geq U^{\mathrm{eq}} \geq & -\frac{1}{2}\int_{*}\mathrm{d}\varepsilon f_{0} .
\end{eqnarray*}
where the symbol $*$ means that the integrals only cover the energy ranges where $\mathrm{Im}(\boldsymbol{G}^r)\neq 0$, or, in other words, the integrals only run over energy ranges where the leads support propagating states, or the system presents localized states.
Then, the above equation shows that the equilibrium potential is bounded by the occupation of the reservoirs and the system.

\section{Equilibrium friction tensor.\label{app:gamma}}
As mentioned, here we are interested only in the symmetric component of the electronic-friction tensor at equilibrium, Eq. \ref{eq:gamma^s}.
Under this condition, the lesser and greater Green's functions are given by~\cite{haug2008,bode2011}:
\begin{eqnarray*}
\boldsymbol{G}_{\mathrm{eq}}^{<} & = & if_{0}\boldsymbol{A}\\
\boldsymbol{G}_{\mathrm{eq}}^{>} & = & -i\left[1-f_{0}\right]\boldsymbol{A}
\end{eqnarray*}
where $\boldsymbol{A}$ is the spectral function, $\boldsymbol{A}=i\left(\boldsymbol{G}^{r}-\boldsymbol{G}^{a}\right)$. Using this we can write Eq. \ref{eq:gamma^s} as 
\begin{eqnarray}
\gamma_{\nu\nu'}^{\mathrm{eq}}  =
\int\frac{\hbar d\varepsilon}{4\pi}f_{0}
\left (
\partial_{\varepsilon}\left(1-f_{0}\right)
\mathbf{\textrm{Tr}} \left [ \left\{ \boldsymbol{\Lambda}_{\nu}\boldsymbol{A}\boldsymbol{\Lambda}_{\nu'}\boldsymbol{A}\right\} _{s}
\right ] \right . & &
\notag \\
 \left .
 + \left(1-f_{0}\right)\mathbf{\textrm{Tr}}\left[\left\{ \boldsymbol{\Lambda}_{\nu}\boldsymbol{A}\boldsymbol{\Lambda}_{\nu'}\boldsymbol{A}\right\} _{s}\partial_{\varepsilon}\boldsymbol{A}\right]
 \right ) \qquad
 & &  \label{eq:gamma_ini}
\end{eqnarray}
where we used the notation
$\left\{ \boldsymbol{\Lambda}_{\nu}\boldsymbol{A}\boldsymbol{\Lambda}_{\nu'}\boldsymbol{A}\right\} _{s} =\left(\boldsymbol{\Lambda}_{\nu}\boldsymbol{A}\boldsymbol{\Lambda}_{\nu'}\boldsymbol{A}+\boldsymbol{\Lambda}_{\nu'}\boldsymbol{A}\boldsymbol{\Lambda}_{\nu}\boldsymbol{A}\right)$.

Employing the cyclic property of the trace of matrix multiplications, one can readily show
\begin{eqnarray}
\mathbf{\textrm{Tr}}
\left[\left\{ \boldsymbol{\Lambda}_{\nu}\boldsymbol{A}\boldsymbol{\Lambda}_{\nu'}\boldsymbol{A}\right\} _{s}\right]
&=&
2\mathbf{\textrm{Tr}}
\left[\boldsymbol{\Lambda}_{\nu}\boldsymbol{A}\boldsymbol{\Lambda}_{\nu'}\boldsymbol{A}\right].
\label{eq:prop1}
\end{eqnarray}
The above equation can be used to further prove
\begin{eqnarray}
\partial_{\varepsilon}\mathbf{\textrm{Tr}}
\left[\boldsymbol{\Lambda}_{\nu}\boldsymbol{A}\boldsymbol{\Lambda}_{\nu'}\boldsymbol{A}\right]
&=&
\mathbf{\textrm{Tr}}
\left[
\left\{\boldsymbol{\Lambda}_{\nu}\boldsymbol{A}\boldsymbol{\Lambda}_{\nu'}\right\}_s
\partial_{\varepsilon}\boldsymbol{A}\right].
\label{eq:prop2}
\end{eqnarray}
Combining Eqs. \ref{eq:prop1} and \ref{eq:prop2} we find
\begin{eqnarray}
2\partial_{\varepsilon}\left(\left(1-f_{0}\right)
\mathbf{\textrm{Tr}}
\left[\boldsymbol{\Lambda}_{\nu}\boldsymbol{A}\boldsymbol{\Lambda}_{\nu'}\boldsymbol{A}\right]
\right) = & &
\notag \\
\left(1-f_{0}\right)\partial_{\varepsilon}\mathbf{\textrm{Tr}}
\left[\boldsymbol{\Lambda}_{\nu}\boldsymbol{A}\boldsymbol{\Lambda}_{\nu'}\boldsymbol{A}\right]
\notag \\
+\mathbf{\textrm{Tr}} \left[
\left\{\boldsymbol{\Lambda}_{\nu}\boldsymbol{A}\boldsymbol{\Lambda}_{\nu'}\right\}_s
\partial_{\varepsilon} \left(\left(1-f_{0}\right)\boldsymbol{A}\right)\right].
\label{eq:prop3}
\end{eqnarray}
Now, after some algebra and using Eqs. \ref{eq:prop1} and \ref{eq:prop3} the following relation is obtained
\begin{eqnarray}
f_{0}\mathbf{\textrm{Tr}}
\left[
\left\{\boldsymbol{\Lambda}_{\nu}\boldsymbol{A}\boldsymbol{\Lambda}_{\nu'}\right\}_s
\partial_{\varepsilon}\left(\left(1-f_{0}\right)\boldsymbol{A}\right)\right]
= & &
\notag \\ 
2\partial_{\varepsilon}\left(f_{0}\left(1-f_{0}\right) \mathbf{\textrm{Tr}}
\left[\boldsymbol{\Lambda}_{\nu}\boldsymbol{A}\boldsymbol{\Lambda}_{\nu'}\boldsymbol{A}\right]
\right)
\notag \\ 
-2\left(\partial_{\varepsilon}f_{0}\right)\left(1-f_{0}\right) \mathbf{\textrm{Tr}}
\left[\boldsymbol{\Lambda}_{\nu}\boldsymbol{A}\boldsymbol{\Lambda}_{\nu'}\boldsymbol{A}\right]
\notag \\ 
-f_{0}\left(1-f_{0}\right)\partial_{\varepsilon}\mathbf{\textrm{Tr}}
\left[\boldsymbol{\Lambda}_{\nu}\boldsymbol{A}\boldsymbol{\Lambda}_{\nu'}\boldsymbol{A}\right].
\label{eq:prop4}
\end{eqnarray}
Inserting Eq. \ref{eq:prop4} into Eq. \ref{eq:gamma_ini} yields
\begin{eqnarray}
\gamma_{\nu\nu'}^{\mathrm{eq}} & = & \left.\frac{\hbar}{2\pi}f_{0}\left(1-f_{0}\right)\mathbf{\textrm{Tr}}\left[\boldsymbol{\Lambda}_{\nu}\boldsymbol{A}\boldsymbol{\Lambda}_{\nu'}\boldsymbol{A}\right]\right|_{-\infty}^{\infty}
\notag \\  &  &
-\int\frac{\hbar d\varepsilon}{2\pi}\left(\partial_{\varepsilon}f_{0}\right)\left(1-f_{0}\right)\mathbf{\textrm{Tr}}\left[\boldsymbol{\Lambda}_{\nu}\boldsymbol{A}\boldsymbol{\Lambda}_{\nu'}\boldsymbol{A}\right]
\notag \\  &  &
-\int\frac{\hbar d\varepsilon}{4\pi}f_{0}\left(1-f_{0}\right)\partial_{\varepsilon}\mathbf{\textrm{Tr}}\left[\boldsymbol{\Lambda}_{\nu}\boldsymbol{A}\boldsymbol{\Lambda}_{\nu'}\boldsymbol{A}\right]
\label{eq:gamma_casi}
\end{eqnarray}
As $\mathbf{\textrm{Tr}}\left[\boldsymbol{\Lambda}_{\nu}\boldsymbol{A}
\boldsymbol{\Lambda}_{\nu'}\boldsymbol{A}\right]$ is bounded, and 
$ \lim_{\varepsilon \rightarrow\pm\infty} f_{0}\left(1-f_{0}\right) = 0$, then
\begin{eqnarray}
\left.\frac{1}{2\pi}f_{0}\left(1-f_{0}\right)\mathbf{\textrm{Tr}}
\left[\boldsymbol{\Lambda}_{\nu}\boldsymbol{A}\boldsymbol{\Lambda}_{\nu'}\boldsymbol{A}\right]\right|_{-\infty}^{\infty} & = & 0
\label{eq:lim0}
\end{eqnarray}
Using this in Eq. \ref{eq:gamma_casi} and after some algebra we finally find
\begin{eqnarray*}
\gamma_{\nu\nu'}^{\mathrm{eq}} & = &	
\int\frac{\hbar d\varepsilon}{4\pi} \left(-\partial_{\varepsilon}f_{0}\right)
\mathbf{\textrm{Tr}}\left[\boldsymbol{\Lambda}_{\nu}\boldsymbol{A}\boldsymbol{\Lambda}_{\nu'}\boldsymbol{A}\right] .
\end{eqnarray*}

\section{Expansion of $F^{\mathrm{ne}}$.\label{app:Fne}}

Let us consider the contribution of the $\alpha$ lead to the CIF ($F_\alpha$),
\begin{eqnarray*}
F = \sum_\alpha F_\alpha , 
\end{eqnarray*}
and define the compact support function $\varphi_\alpha\left(\varepsilon\right)$ such that
\begin{eqnarray*}
F_\alpha & = & \underset{\mathbb{R}}{\int}\varphi_\alpha\left(\varepsilon\right)f_\alpha \mathrm{d}\varepsilon ,
\end{eqnarray*}
where $f_\alpha$ is the Fermi distribution function of reservoir $\alpha$ and 
\begin{eqnarray}
\varphi_\alpha = \frac{1}{\pi} \mathrm{Tr}
\left [  \boldsymbol{\Lambda}_\alpha \boldsymbol{G}^r \boldsymbol{\Gamma}_\alpha  \boldsymbol{G}^a \right ]
\label{eq:varphi}
\end{eqnarray}
Expanding $F_\alpha$ up to second order in $\mu_\alpha$ and $T_\alpha$ and taking the small temperature limit of $T_\alpha$ yields
\begin{eqnarray}
& F_\alpha & = \stackrel[\infty]{\mu_{0}}{\int}\varphi_\alpha\left(\varepsilon\right)\mathrm{d}\varepsilon
+\frac{\pi^{2}}{6}\left.\frac{\partial\varphi_\alpha}{\partial\varepsilon}\right|_{\varepsilon=\mu_\alpha}\left(k_{B}T_0\right)^{2}
\notag \\
 &  + & 
 \left. \varphi_\alpha \right |_{\varepsilon=\mu_{0}} \delta \mu_\alpha
 +\frac{\pi^{2}}{3}\left.\frac{\partial\varphi_\alpha}{\partial\varepsilon}\right|_{\varepsilon=\mu_0}
 \left(k_{B} T_0 \right)^2 \frac{\delta T_\alpha}{T_0}
 \notag \\
&  + &
\frac{1}{2}\left.\frac{\partial\varphi_\alpha}{\partial\varepsilon}\right|_{\varepsilon=\mu_0} \delta \mu_\alpha^{2}
+\frac{\pi^{2}}{6}\left.\frac{\partial\varphi_\alpha}{\partial\varepsilon}\right|_{\varepsilon=\mu_0}
 k_{B}^2 \delta T_\alpha^{2} .
 \label{eq:F_exp}
 \end{eqnarray}
As in the main text, here we used $\delta T_\alpha = T_\alpha - T_0$ and $\delta \mu_\alpha = \mu_\alpha - \mu_0$, where $T_0$ and $\mu_0$ are the average temperature and chemical potential of the reservoirs connected to the system. Note that the first two terms of the right-hand side of Eq. \ref{eq:F_exp} will contribute to the equilibrium force while the last two terms are second order in an expansion in terms of $\delta \mu_\alpha$ and $\delta T_\alpha$.
Therefore, by inserting Eq. \ref{eq:varphi} into the above equation, taking only the first-order terms and summing up all $F_\alpha$ contributions, one readily arrives at Eq. \ref{eq:F^ne_TVlow}.

\section{Green Functions in our system. \label{app:GFs}}
The determinant of the effective Hamiltonian given in Eq. \ref{eq:H_eff} is
\begin{eqnarray}
\det\left(\varepsilon\boldsymbol{I}-\boldsymbol{H}_{\mathrm{eff}}\right) = V_{0}^{3}\left(\epsilon-\frac{\Sigma_{0}^{r}}{V_{0}}\right)^{2} \times
\notag \\
\left[\epsilon-\epsilon_{d}+i\frac{\Gamma_{\eta}^{r}}{V_{0}}-\left(v_{R}^{2}+v_{L}^{2}\right)\frac{\Sigma_{0}^{r}}{V_{0}}\right] .
\label{eq:det[H_eff]}
\end{eqnarray}
The retarded Green function is
\begin{eqnarray*}
\boldsymbol{G}^{r}\left(\varepsilon\right) & =
\begin{pmatrix}G_{11}^{r} & G_{12}^{r} & G_{13}^{r}\\
G_{21}^{r} & G_{22}^{r} & G_{23}^{r}\\
G_{31}^{r} & G_{32}^{r} & G_{33}^{r}
\end{pmatrix}
& = \left [ \epsilon \boldsymbol{I} - \boldsymbol{H}_{\mathrm{eff}}\right ]^{-1}
. \notag \label{eq:G^r_definition}
\end{eqnarray*}
For convenience, we changed the notation for the indexes of the elements of $\boldsymbol{G}$, with respect to that of Eq. \ref{eq:H_TB} in the main text.
Thus, e.g., $-(1/\pi) \mathrm{Im}(G_{2,2}^r)$ is the LDOS of the dot (site "$0$" according to Eq. \ref{eq:H_TB}), while $-(1/\pi) \mathrm{Im}(G_{1,1}^r)$ is the LDOS of the first site of the left chain (site "$-1$" according to Eq. \ref{eq:H_TB}). In all the appendixes, we followed the present convention.

Using Eq. \ref{eq:det[H_eff]} one can readily calculate the elements of $\boldsymbol{G}^{r}\left(\varepsilon\right)$.
Written them in terms of the dimensionless quantities ($\epsilon$, $\epsilon_{d}$, $v_{L/R}$, $\widetilde{\Gamma}_{\eta}=\frac{\Gamma_{\eta}}{V_{0}}$, and $\widetilde{\Sigma}_{0}^{r}=\frac{\Sigma_{0}^{r}\left(\varepsilon\right)}{V_{0}}$) they are
\begin{eqnarray*}
G_{11}^{r} & = & \frac{\left(\epsilon-\epsilon_{d}+i\widetilde{\Gamma}_{\eta} -v_{R}^{2}\widetilde{\Sigma}_{0}^{r}\right)}{V_{0}
\left[\epsilon-\epsilon_{d}+i\widetilde{\Gamma}_{\eta}-\left(v_{R}^{2}+v_{L}^{2}\right)\widetilde{\Sigma}_{0}^{r}\right]
\left(\epsilon-\widetilde{\Sigma}_{0}^{r}\right)}\\
G_{22}^{r} & = & \frac{1}{V_{0}
\left[\epsilon-\epsilon_{d}+i\widetilde{\Gamma}_{\eta}-\left(v_{R}^{2}+v_{L}^{2}\right)\widetilde{\Sigma}_{0}^{r}\right]
}\\
G_{33}^{r} & = & \frac{\left(\epsilon-\epsilon_{d}+i\widetilde{\Gamma}_{\eta}-v_{L}^{2}\widetilde{\Sigma}_{0}^{r}\right)}{V_{0}
\left[\epsilon-\epsilon_{d}+\widetilde{\Gamma}_{\eta}-\left(v_{R}^{2}+v_{L}^{2}\right)\widetilde{\Sigma}_{0}^{r}\right] 
\left(\epsilon-\widetilde{\Sigma}_{0}^{r}\right)}\\
G_{21}^{r} & = & G_{12}^{r}=\frac{-v_{L}\widetilde{\Sigma}_{0}^{r}}{V_{0}
\left[\epsilon-\epsilon_{d}+i\widetilde{\Gamma}_{\eta}-\left(v_{R}^{2}+v_{L}^{2}\right)\widetilde{\Sigma}_{0}^{r}\right]
}\\
G_{32}^{r} & = & G_{23}^{r}=\frac{-v_{R}\widetilde{\Sigma}_{0}^{r}}{V_{0}
\left[\epsilon-\epsilon_{d}+i\widetilde{\Gamma}_{\eta}-\left(v_{R}^{2}+v_{L}^{2}\right)\widetilde{\Sigma}_{0}^{r}\right]
}\\
G_{31}^{r} & = & G_{13}^{r}=\frac{v_{L}v_{R}\left(\widetilde{\Sigma}_{0}^{r}\right)^{2}}{V_{0}
\left[\epsilon-\epsilon_{d}+i\widetilde{\Gamma}_{\eta}-\left(v_{R}^{2}+v_{L}^{2}\right)\widetilde{\Sigma}_{0}^{r}\right]
}
\end{eqnarray*}
The advanced Green function is calculated from $\boldsymbol{G}^{a}=\left[\boldsymbol{G}^{r}\right]^{\dagger}$, while the lesser Green function is given by
\begin{equation*}
\boldsymbol{G}^{<}=\boldsymbol{G}^{r}\boldsymbol{\Sigma}^{<}\boldsymbol{G}^{a},
\end{equation*}
where $\boldsymbol{\Sigma}^{<}=\boldsymbol{\Sigma}^{<}_L+\boldsymbol{\Sigma}^{<}_R+\boldsymbol{\Sigma}^{<}_\eta$. 
We will split $\boldsymbol{G}^{<}$ into two contributions ($\boldsymbol{G}_{(L+R)}^{<}$ and $\boldsymbol{G}_{\eta}$)
\begin{eqnarray*}
\boldsymbol{G}^{<} & = & \underbrace{\boldsymbol{G}^{r}\left(\boldsymbol{\Sigma}_{L}^{<}
+\boldsymbol{\Sigma}_{R}^{<}\right)\boldsymbol{G}^{a}}_{\boldsymbol{G}_{(L+R)}^{<}}+
\underbrace{\boldsymbol{G}^{r}\boldsymbol{\Sigma}_{\eta}^{<}\boldsymbol{G}^{a}}
_{\boldsymbol{G}_{\eta}^{<}}
\end{eqnarray*}
where
\begin{eqnarray*}
\left(\boldsymbol{\Sigma}_{L}^{<}+\boldsymbol{\Sigma}_{R}^{<}\right) & = & 2i\begin{pmatrix}\left(f_{0}+\frac{\Delta f}{2}\right)\Gamma_{L} & 0 & 0\\
0 & 0 & 0\\
0 & 0 & \left(f_{0}-\frac{\Delta f}{2}\right)\Gamma_{R}
\end{pmatrix}\\
\boldsymbol{\Sigma}_{\eta}^{<} & = & 2i\begin{pmatrix}0 & 0 & 0\\
0 & f_{0}\Gamma_{\eta}^{\left(0\right)} & 0\\
0 & 0 & 0
\end{pmatrix}.
\end{eqnarray*}
Here, $f_{L/R}=f_0\pm \Delta f$ with $f_0=(f_L+f_R)/2$ and $\Delta f = f_L-f_R$. We also assumed the third lead is at equilibrium with $f_{\eta}=f_{0}$.
Then, the elements of $\boldsymbol{G}_{(L+R)}^{<}$ and $\boldsymbol{G}_{\eta}^{<}$
are
\begin{eqnarray*}
\left[G_{(L+R)}^{<}\right]_{ij} & = & \underbrace{2if_{0}\left(G^r_{i1}G^{r*}_{j1}\Gamma_{L}+G^{r}_{i3}G^{r*}_{j3}\Gamma_{R}\right)}_{\left[G_{(L+R)}^{<}\right]_{ij}^{\mathrm{(eq)}}}\\
 &  &  +\underbrace{2i\frac{\Delta f}{2}\left(G^{r}_{i1}G^{r*}_{j1}\Gamma_{L}-G^{r}_{i3}G^{r*}_{j3}\Gamma_{R}\right)}_{\left[G_{(L+R)}^{<}\right]_{ij}^{\mathrm{(ne)}}}
 \end{eqnarray*}
 and
 \begin{eqnarray*}
\left[G_{\eta}^{<}\right]_{ij} & = & 2if_{0}\left(G^{r}_{i2}G^{r*}_{j2}\Gamma_{\eta}^{\left(0\right)}\right).
\end{eqnarray*}
where we identified the equilibrium (eq) and nonequilibrium (ne) contributions to $\left[\boldsymbol{G}_{(L+R)}^{<}\right]_{ij}$. The equilibrium contributions only contains $f_{0}$ terms while nonequilibrium contributions only contains $\Delta f$ terms.

Therefore, the nonequilibrium $\boldsymbol{G}^{<}$ is given by
\begin{eqnarray*}
\boldsymbol{G}_{(\mathrm{ne})}^{<} & = & \left[\boldsymbol{G}_{(L+R)}^{<}\right]_{ij}^{\mathrm{(ne)}},
\end{eqnarray*}
while the equilibrium contribution is
\begin{eqnarray*}
\boldsymbol{G}_{(\mathrm{eq})}^{<} & = &
\begin{cases}
\left[\boldsymbol{G}_{(L+R)}^{<}\right]^{(\mathrm{eq})} & \left|\epsilon\right|\leq2\\
\lim_{\Gamma_{\eta}\rightarrow0} \boldsymbol{G}_{\eta}^{<}  & \left|\epsilon\right|>2
\end{cases}.
\end{eqnarray*}

\section{Mechanical variables and CIFs. \label{app:mech_var}}
Let $\left\{  q_{j}\right\}$ and $\left\{  X_{\nu}\right\}$
be two sets of different mechanical variables related through the expression $q_{j}\left(X_{\nu}\right)$, which we assume continuous but not necessarily lineal. 
Then, CIFs can be written in terms of $\left\{  X_{\nu} \right\}$ as $F\left(X_{\nu}\right)$ or in terms of $\left\{  q_{j}\right\}$ as $F\left(q_{j}\right)$. Both descriptions are related through
\begin{eqnarray*}
F\left(X_{\nu}\right) & = & \int\frac{\mathrm{d}\varepsilon}{2\pi i}\mathbf{\textrm{Tr}}\left[\boldsymbol{\Lambda}_{\nu}\boldsymbol{G}^{<}\right]
\notag \\ & = &
\underset{j}{\sum}
\underbrace{
\left(\int\frac{\mathrm{d}\varepsilon}{2\pi i}\mathbf{\textrm{Tr}}\left[\boldsymbol{\Lambda}_{j}\boldsymbol{G}^{<}\right]\right)
}_{F\left(q_{j}\right)}
\frac{\partial q_{j}}{\partial X_{\nu}} .
\end{eqnarray*}
where $\frac{\partial q_{j}}{\partial X_{\nu}}$ is the covariant derivative.

The work done by the forces over a closed loop $C$ in the space of the variables is independent of the choice of the sets of mechanical variables used to describe the system, as can be readily shown
\begin{eqnarray*}
W\left(\overrightarrow{X}\right) & = &
\underset{C\left(\overrightarrow{X}\right)}{\int}\underset{\nu}{\sum}F\left(X_{\nu}\right)\mathrm{d}X_{\nu}
\notag \\ & = &
\underset{\nu}{\sum}\underset{j}{\sum}\underset{C\left(\overrightarrow{q}\right)}{\int}F\left(q_{j}\right)\frac{\partial q_{j}}{\partial X_{\nu}}\mathrm{d}X_{\nu}
\notag \\ & = &
\underset{C\left(\overrightarrow{q}\right)}{\int}\underset{j}{\sum} F\left(q_{j}\right)\mathrm{d}q_{j} =W\left(\overrightarrow{q}\right)
\end{eqnarray*}
In the main text, we used this property to choose a set of convenient variables to become independent of the particular way in which mechanical variables affect the electronic Hamiltonian.

\section{General expression of the curl \label{app:curl}}

Our goal here is to obtained a closed-form expression for $\left(\nabla\times F\right)_{\rho}$.
We start by evaluating the derivative of the force $\partial_{\nu'}F_{\nu}=\partial F_{\nu}/\partial X_{\nu'}$
\begin{eqnarray*}
\partial_{\nu'}F_{\nu} & = & \frac{1}{2\pi i}\stackrel[-\infty]{\infty}{\int}\partial_{\nu'}\left\{ \textrm{Tr}\left(\boldsymbol{\Lambda}_{\nu}\boldsymbol{G}^{<}\right)\right\} d\varepsilon.
\end{eqnarray*}
The derivative of the lesser Green function is
\begin{eqnarray*}
\partial_{\nu'}\boldsymbol{G}^{<} & = &
\boldsymbol{G}^{r}\boldsymbol{\Lambda}_{\nu'}\boldsymbol{G}^{<}
+\boldsymbol{G}^{<}\boldsymbol{\Lambda}_{\nu'}G^{a}
\end{eqnarray*}
where we assume $\partial_{\nu'}\boldsymbol{\Sigma}^{<}=0$ and used $\partial_{\nu'}\boldsymbol{G}^{r/a} = \boldsymbol{G}^{r/a}\boldsymbol{\Lambda}_{\nu'}\boldsymbol{G}^{r/a}$.
Then, the derivative of the trace is 
\[
\begin{array} {lcc}
\partial_{\nu'}\textrm{Tr}\left(\boldsymbol{\Lambda}_{\nu}\boldsymbol{G}^{<}\right) = & & \\
\textrm{Tr}\left(\left[\partial_{\nu'}\boldsymbol{\Lambda}_{\nu}
-\boldsymbol{\Lambda}_{\nu}\boldsymbol{G}^{r}\boldsymbol{\Lambda}_{\nu'}
-\left(\boldsymbol{\Lambda}_{\nu}\boldsymbol{G}^{r}\boldsymbol{\Lambda}_{\nu'}\right)^{\dagger}
\right]\boldsymbol{G}^{<}\right)
\end{array}
\]
where we used $\boldsymbol{\Lambda}_{\nu'}^{\dagger}=\boldsymbol{\Lambda}_{\nu'}$
and $\left[\boldsymbol{G}^{r}\right]^{\dagger}=\boldsymbol{G}^{a}$
to make $\boldsymbol{\Lambda}_{\nu'}G^{a}\boldsymbol{\Lambda}_{\nu}=\left[\boldsymbol{\Lambda}_{\nu}\boldsymbol{G}^{r}\boldsymbol{\Lambda}_{\nu'}\right]^{\dagger}$.
Using this, the derivative of the force can be written as
\begin{eqnarray*}
\partial_{\nu'}F_{\nu} & = &
\frac{1}{2\pi i}\stackrel[-\infty]{\infty}{\int}\textrm{Tr}\left(\left[\boldsymbol{\Lambda}_{\nu} \boldsymbol{G}^{r} \boldsymbol{\Lambda}_{\nu'}^{}
\right . \right . \notag \\ &&
\left . \left .
+\left(\boldsymbol{\Lambda}_{\nu} \boldsymbol{G}^{r} \boldsymbol{\Lambda}_{\nu'}\right)^{\dagger}
-\partial_{\nu'}\boldsymbol{\Lambda}_{\nu} \right] \boldsymbol{G}^{<}\right)d\varepsilon.
\end{eqnarray*}
Now as $\partial_{\nu'} \boldsymbol{\Lambda}_{\nu}=\partial_{\nu} \boldsymbol{\Lambda}_{\nu'}$,
the following holds
\begin{eqnarray*}
\partial_{\nu'}F_{\nu}-\partial_{\nu}F_{\nu'} & = &
\frac{1}{2\pi i}\stackrel[-\infty]{\infty}{\int}\textrm{Tr}\left(
\left[\boldsymbol{\Pi}_{\nu'\nu} -\boldsymbol{\Pi}_{\nu\nu'}\right] \boldsymbol{G}^{<}\right)d\varepsilon
\end{eqnarray*}
where, to compact the notation, we define
\begin{eqnarray*}
\boldsymbol{\Pi}_{\nu'\nu} & = & \boldsymbol{\Lambda}_{\nu'}\boldsymbol{G}^{r} \boldsymbol{\Lambda}_{\nu}+
\left(\boldsymbol{\Lambda}_{\nu'}\boldsymbol{G}^{r}\boldsymbol{\Lambda}_{\nu}\right)^{\dagger}.
\end{eqnarray*}
Finally, the curl of the force yields
\begin{eqnarray*}
\left(\nabla\times\overrightarrow{F}\right)_{\rho} & = &
\frac{1}{2\pi i}\stackrel[-\infty]{\infty}{\int}
\textrm{Tr}\left( \epsilon^{\rho\nu'\nu}   \boldsymbol{\Pi}_{\nu'\nu} \boldsymbol{G}^{<}          \right) d\varepsilon,
\end{eqnarray*}
where $\epsilon^{\rho\mu\nu}$ is the Levi-Civita symbol.

\section{Elements of the curl in the TB system.\label{app:curl_TB}}

We start by defining the index order of the variables
\begin{eqnarray*}
E_{d} & \rightarrow & \#1\\
V_{L} & \rightarrow & \#2\\
V_{R} & \rightarrow & \#3 .
\end{eqnarray*}
With this order, the following relations hold
\begin{eqnarray*}
\epsilon^{E_{d}\mu\nu}\boldsymbol{\Pi}_{\mu\nu} & = &
\left \{
\boldsymbol{\Lambda}_{V_{R}}\left(\boldsymbol{G}^{r}-\boldsymbol{G}^{a}\right)\boldsymbol{\Lambda}_{V_{L}}
\right \}_a
\\
\epsilon^{V_{L}\mu\nu}\boldsymbol{\Pi}_{\mu\nu} & = &
\left \{
\boldsymbol{\Lambda}_{E_{d}}\left(\boldsymbol{G}^{r}-\boldsymbol{G}^{a}\right)\boldsymbol{\Lambda}_{V_{R}}
\right \}_a
\\
\epsilon^{V_{R}\mu\nu}\boldsymbol{\Pi}_{\mu\nu} & = &
\left \{
\boldsymbol{\Lambda}_{V_{L}}\left(\boldsymbol{G}^{r}-\boldsymbol{G}^{a}\right)\boldsymbol{\Lambda}_{E_{d}}
\right \}_a
\end{eqnarray*}
where
$\left \{ \boldsymbol{\lambda}_\nu...\boldsymbol{\lambda}_{\nu'} \right \}_a =
\left (\boldsymbol{\lambda}_\nu...\boldsymbol{\lambda}_{\nu'} 
-\boldsymbol{\lambda}_{\nu'}...\boldsymbol{\lambda}_{\nu}\right )$.

Defining $\Sigma_{L/R}= v_{L/R}^{2}\Sigma_{0}^{r}$ and $\Sigma_{L/R}=\Delta_{L/R}-i\Gamma_{L/R}$, one can deduce the following
\begin{eqnarray}
\frac{1}{V_{L/R}^{2}}\left|\Sigma_{L/R}\right|^{2}\Gamma_{0} & = &
\frac{V_{L/R}^{2}\Gamma_{0}}{\left|\varepsilon-\left(E_{0}+\Sigma_{0}^{r}\right)\right|^{2}}
\label{eq:useful_Sig^2}
\\
 & = & \Gamma_{L/R}. 
\end{eqnarray}
This relation will be useful afterward. Another formula that will
be useful is the expression for the elements of $G^{<}$. For simplicity,
we will assume $\Gamma_{\eta}=0$ from the beginning, as we are interested only in the nonequilirbium contribution of the force, see appendix \ref{app:GFs}.
Then, the lesser self-energy yields
\begin{eqnarray*}
\boldsymbol{\Sigma}^{<}  & = & 2i\begin{pmatrix}f_{L}\Gamma_{L} & 0 & 0\\
0 & 0 & 0\\
0 & 0 & f_{R}\Gamma_{R}
\end{pmatrix}.
\end{eqnarray*}
It will be helpful to define $\boldsymbol{G}^{\lesssim}=\frac{1}{2i}\boldsymbol{G}^{<}$. In our case, the elements of $\boldsymbol{G}^{\lesssim}$ are
\begin{eqnarray}
G_{ij}^{\lesssim} & = & \left\{ G^{r}_{i1}G^{r*}_{j1}f_{L}\Gamma_{L}+G^{r}_{i3}G^{r*}_{j3}f_{R}\Gamma_{R}\right\} \label{eq:G^=D}
\end{eqnarray}
Note that $G_{ij}^{\lesssim*}=G_{ji}^{\lesssim}$. 
In the following, we will use all the above properties and definitions to derive the expression for each of the elements
of the curl.

\subsection{Element $\rho=E_{d}$.\label{app:rho=E_d}}

This element of the curl is given by
\begin{eqnarray*}
\left(\nabla\times\overrightarrow{F}\right)_{E_{d}} & = & \frac{1}{2\pi i}\stackrel[-\infty]{\infty}{\int}\mathbf{\textrm{Tr}}\left[\left(\epsilon^{E_{d}\mu\nu}\boldsymbol{\Pi}_{\mu\nu}\right)
\boldsymbol{G}^{<}\right]
\end{eqnarray*}
where
\begin{eqnarray*}
\epsilon^{E_{d}\mu\nu}\boldsymbol{\Pi}_{\mu\nu} & = &
\left \{
\boldsymbol{\Lambda}_{V_{R}} \left( \boldsymbol{G}^{r}-\boldsymbol{G}^{a}\right) \boldsymbol{\Lambda}_{V_{L}}
\right \}_a
\end{eqnarray*}
Now, the terms of the right-hand side of the above equation are
\begin{eqnarray*}
\boldsymbol{\Lambda}_{V_{R}} \left(\boldsymbol{G}^{r}-\boldsymbol{G}^{a}\right)
\boldsymbol{\Lambda}_{V_{L}} & = & \begin{pmatrix}0 & 0 & 0\\
a_{32} & a_{31} & 0\\
a_{22} & a_{21} & 0
\end{pmatrix}
\end{eqnarray*}
\begin{eqnarray*}
\boldsymbol{\Lambda}_{V_{L}}\left( \boldsymbol{G}^{r}-\boldsymbol{G}^{a}\right)
\boldsymbol{\Lambda}_{V_{R}} & = & \begin{pmatrix}0 &
a_{23} & a_{22}\\
0 & a_{13} & a_{12}\\
0 & 0 & 0
\end{pmatrix},
\end{eqnarray*}
and therefore 
\begin{eqnarray*}
\epsilon^{E_{d}\mu\nu}\boldsymbol{\Pi}_{\mu\nu} & = &
\begin{pmatrix}0 & -a_{23} & -a_{22}\\
a_{32} & a_{31}-a_{13} & -a_{12}\\
a_{22} & a_{21} & 0
\end{pmatrix} .
\end{eqnarray*}
Here, we define $a_{ij}=\left(G^{r}_{ij}-G^{r*}_{ji}\right)$ to compact the notation.
As the retarded Green functions is symmetric in our case,
the following relations holds $a_{ij}=2i \mathrm{Im}\{G^{r}_{ij}\}$, $a_{31} = a_{13}$, $a_{32} = a_{23}$, and $a_{21} = a_{12}$.
Using this, we obtain
\begin{eqnarray}
\mathbf{\textrm{Tr}}\left[\left(\epsilon^{E_{d}\mu\nu}\boldsymbol{\Pi}_{\mu\nu}\right)
\boldsymbol{G}^{<}\right] =
& & \notag \\
2i
\left\{
a_{23}b_{12}
+a_{22}b_{13}
+a_{12}b_{23}
\right\} 
& &
\label{eq:Tr=2iab}
\end{eqnarray}
where we defined $b_{ij}=\left(G_{ij}^{\lesssim}-G_{ij}^{\lesssim *}\right)=2i \mathrm{Im}\{G_{ij}^{\lesssim}\}$ with $G_{ij}^{\lesssim}$ given by Eq. \ref{eq:G^=D}.
After some algebra, the right-hand side terms of the above equation can be written
as
\begin{eqnarray*}
a_{23}b_{12} =
-4\textrm{Im}\left\{ G^{r}_{23}\right\} \times
\\
\left(\textrm{Im}\left\{ G^{r}_{11}G^{r*}_{21}\right\} f_{L}\Gamma_{L}+\textrm{Im}\left\{ G^{r}_{13}G^{r*}_{23}\right\} f_{R}\Gamma_{R}\right)
\end{eqnarray*}
\begin{eqnarray*}
a_{22}b_{13} = -4\textrm{Im}\left\{ G^{r}_{22}\right\} \times
\\ 
\left(\textrm{Im}\left\{ G^{r}_{11}G^{r*}_{31}\right\} f_{L}\Gamma_{L}+\textrm{Im}\left\{ G^{r}_{13}G^{r*}_{33}\right\} f_{R}\Gamma_{R}\right)
\end{eqnarray*}
\begin{eqnarray*}
a_{12}b_{23} = -4\textrm{Im}\left\{ G^{r}_{12}\right\} \times \\
\left(\textrm{Im}\left\{ G^{r}_{21} G^{r *}_{31}\right\} f_{L}\Gamma_{L}+\textrm{Im}\left\{ G^{r}_{23}G^{r *}_{33}\right\} f_{R}\Gamma_{R}\right)
\end{eqnarray*}

Using the expression for $\boldsymbol{G}^{r}$ given in appendix \ref{app:GFs} and Eq. \ref{eq:useful_Sig^2} we finds
\begin{eqnarray}
\textrm{Im}\left\{ G^{r}_{11}G^{r*}_{21}\right\} f_{L}\Gamma_{L}+\textrm{Im}\left\{ G^{r}_{13}G^{r*}_{23}\right\} f_{R}\Gamma_{R} & = & \notag \\
\frac{\Gamma_{L}\Gamma_{R}}{\left|\varepsilon-E_{d}-\Sigma_{L}-\Sigma_{R}\right|^{2}}\frac{\left[f_{R}-f_{L}\right]}{V_{L}}\label{eq:Im(GG)-01} ,
\end{eqnarray}
\begin{eqnarray}
\textrm{Im}\left\{ G^{r}_{11}G^{r*}_{31}\right\} f_{L}\Gamma_{L}+\textrm{Im}\left\{ G^{r}_{13}G^{r*}_{33}\right\} f_{R}\Gamma_{R} & = & \notag \\
-\frac{\Gamma_{L}\Gamma_{R}}{\left|\varepsilon-E_{d}-\Sigma_{L}-\Sigma_{R}\right|^{2}}\frac{\left(\varepsilon-E_{d}\right)}{V_{L}V_{R}}\left[f_{R}-f_{L}\right]\label{eq:Im(GG)-02}
\end{eqnarray}
and
\begin{eqnarray}
\textrm{Im}\left\{ G^{r}_{21}G^{r*}_{31}\right\} f_{L}\Gamma_{L}+\textrm{Im}\left\{ G^{r}_{23}G^{r*}_{33}\right\} f_{R}\Gamma_{R} & = &  \notag \\
\frac{\Gamma_{L}\Gamma_{R}}{\left|\varepsilon-E_{d}-\Sigma_{L}-\Sigma_{R}\right|^{2}}\frac{\left[f_{R}-f_{L}\right]}{V_{R}}\label{eq:Im(GG)-03}.
\end{eqnarray}

Now introducing the transmittance $\widetilde{T}\left(\varepsilon\right)$
~\cite{haug2008,pastawski2001}
\begin{equation}
\begin{array}{ccc}
\widetilde{T}\left(\varepsilon\right) & = & 2\Gamma_{L}\left|G^{r}_{22}\right|^{2}2\Gamma_{R}\\
 & = & 4\Gamma_{L}\frac{1}{\left|\varepsilon-E_{d}-\Sigma_{L}-\Sigma_{R}\right|^{2}}\Gamma_{R}
\end{array}\label{eq:T_TB}
\end{equation}
and using Eqs. \ref{eq:Im(GG)-01}, \ref{eq:Im(GG)-02}, and \ref{eq:Im(GG)-03} into Eq. \ref{eq:Tr=2iab} gives
\begin{eqnarray*}
\mathbf{\textrm{Tr}}\left[\left(\epsilon^{E_{d}\mu\nu} \boldsymbol{\Pi}_{\mu\nu}\right)
\boldsymbol{G}^{<}\right] = -2i\left\{ \frac{\textrm{Im}\left\{ G^{r}_{23}\right\}}{V_{L}} \times
\right . \notag \\  
\left .
- \frac{\textrm{Im}\left\{ G^{r}_{22}\right\}}{V_{L}V_{R}}\left(\varepsilon-E_{d}\right) 
+ \frac{\textrm{Im}\left\{ G^{r}_{12}\right\}}{V_{R}}\right\} \widetilde{T}\left(\varepsilon\right)\left[f_{R}-f_{L}\right]
\end{eqnarray*}

Using the expressions for the elements of $\boldsymbol{G}^{r}$, it is not difficult to prove:
\begin{eqnarray*}
\textrm{Im}\left\{ G^{r}_{23}\right\} \frac{1}{V_{L}}+\textrm{Im}\left\{ G^{r}_{12}\right\} \frac{1}{V_{R}} & = & -\frac{\left(\varepsilon-E_{d}\right)}{V_{L}V_{R}}\textrm{Im}\left\{ G^{r}_{22}\right\} 
\end{eqnarray*}
Therefore, the element $\rho=E_{d}$ of the curl can be written as
\begin{eqnarray*}
\left(\nabla\times\overrightarrow{F}\right)_{E_{d}} & = & -2\stackrel[-\infty]{\infty}{\int}\frac{\left(\varepsilon-E_{d}\right)}{V_{L}V_{R}}N_{d}\widetilde{T}\left(\varepsilon\right)\left[f_{R}-f_{L}\right]d\varepsilon
\end{eqnarray*}
where $N_{d}$ is the LDOS of the dot
\begin{equation}
\begin{array}{ccc}
N_{d} & = & -\frac{1}{\pi}\underset{\eta\rightarrow0^{+}}{\lim}\textrm{Im}\left\{ G^{r}_{22}\left(\varepsilon+i\eta\right)\right\} \end{array}\label{eq:Nd_TB}
\end{equation}

\subsection{Element $\rho=V_{L}$.\label{app:rho=V_L}}

This element of the curl is given by
\begin{eqnarray*}
\left(\nabla_{\overrightarrow{X}}\times\overrightarrow{F}\right)_{V_{L}} & = & \frac{1}{2\pi i}\stackrel[-\infty]{\infty}{\int}\mathbf{\textrm{Tr}}\left[\left(\epsilon^{V_{L}\mu\nu} \boldsymbol{\Pi}_{\mu\nu}\right) \boldsymbol{G}^{<}\right]
\end{eqnarray*}
where
\begin{eqnarray*}
\epsilon^{V_{L}\mu\nu}\boldsymbol{\Pi}_{\mu\nu} & = &
\left \{
\boldsymbol{\Lambda}_{E_{d}} \left(\boldsymbol{G}^{r}-\boldsymbol{G}^{a} \right)\boldsymbol{\Lambda}_{V_{R}}
\right \}_a
\end{eqnarray*}
Now, the terms of the right-hand side of the above equation are
\begin{eqnarray*}
\boldsymbol{\Lambda}_{E_{d}}\left( \boldsymbol{G}^{r}-\boldsymbol{G}^{a}\right) \boldsymbol{\Lambda}_{V_{R}} & = & -\begin{pmatrix}0 & 0 & 0\\
0 & a_{23} & a_{22}\\
0 & 0 & 0
\end{pmatrix}
\end{eqnarray*}
\begin{eqnarray*}
\boldsymbol{\Lambda}_{V_{R}} \left(\boldsymbol{G}^{r}-\boldsymbol{G}^{a}\right) \boldsymbol{\Lambda}_{E_{d}} & = & -\begin{pmatrix}0 & 0 & 0\\
0 & a_{32} & 0\\
0 & a_{22} & 0
\end{pmatrix}
\end{eqnarray*}
Therefore
\begin{eqnarray*}
\epsilon^{V_{L}\mu\nu}\boldsymbol{\Pi}_{\mu\nu} & = &
\begin{pmatrix}0 & 0 & 0\\
0 & a_{32}-a_{23} & -a_{22}\\
0 & a_{22} & 0
\end{pmatrix}
\end{eqnarray*}
Using this we obtain
\begin{eqnarray*}
\mathbf{\textrm{Tr}}\left[\left(\epsilon^{V_{L}\mu\nu}\boldsymbol{\Pi}_{\mu\nu}\right) \boldsymbol{G}^{<}\right] & = & 2ia_{22}b_{23} \label{eq:TrVLD}
\end{eqnarray*}
where the elements of the matrix $G^{\lesssim}$ are given in Eq. \ref{eq:G^=D}.
After some algebra it is not difficult to show 
\[
\begin{array} {lcc}
a_{22}b_{23} =  -4\textrm{Im}\left\{ G^{r}_{22}\right\} \times
\notag \\ 
\left(\textrm{Im}\left\{ G^{r}_{21}G^{r*}_{31}\right\} f_{L}\Gamma_{L}+\textrm{Im}\left\{ G^{r}_{23}G^{r*}_{33}\right\} f_{R}\Gamma_{R}\right)
\end{array}
\]
Using the expression for $\boldsymbol{G}^{r}$ given in appendix \ref{app:GFs} and Eq. \ref{eq:useful_Sig^2} one finds 
\begin{eqnarray}
\textrm{Im}\left\{ G^{r}_{21}G^{r*}_{31}\right\} f_{L}\Gamma_{L}+\textrm{Im}\left\{ G^{r}_{23}G^{r*}_{33}\right\} f_{R}\Gamma_{R} & = &
\notag \\ 
\frac{\Gamma_{L}\Gamma_{R}}{\left|\varepsilon-E_{d}-\Sigma_{L}-\Sigma_{R}\right|^{2}}\frac{1}{V_{R}}\left[f_{R}-f_{L}\right]\label{eq:Im(GG)-04}
\end{eqnarray}
Using Eqs. \ref{eq:T_TB}, \ref{eq:Nd_TB}, \ref{eq:Im(GG)-04} into Eq. \ref{eq:TrVLD}, we obtain
\begin{eqnarray*}
\mathbf{\textrm{Tr}}\left[\left(\epsilon^{V_{L}\mu\nu}\Pi_{\mu\nu}\right)G^{<}\right] & = & -2\pi i\frac{1}{V_{R}}N_{d}\widetilde{T}\left(\varepsilon\right)\left[f_{R}-f_{L}\right].
\end{eqnarray*}
Therefore, the element $\rho=V_{L}$ of the curl can be written as 
\begin{eqnarray*}
\left(\nabla\times\overrightarrow{F}\right)_{V_{L}} & = & -\frac{1}{V_{R}}\stackrel[-\infty]{\infty}{\int}N_{d}\widetilde{T}\left(\varepsilon\right)\left[f_{R}-f_{L}\right]d\varepsilon
\end{eqnarray*}

\subsection{Element $\rho=V_{R}$.\label{app:rho=V_R}}

The expression for the element $\rho=V_{R}$ of the curl is obtained by following a similar procedure to that of $\left(\nabla\times\overrightarrow{F}\right)_{V_{L}}$ but replacing $V_L$ by $V_R$.
The result is
\begin{eqnarray*}
\left(\nabla\times\overrightarrow{F}\right)_{V_{R}} & = & -\frac{1}{V_{L}}\stackrel[-\infty]{\infty}{\int}N_{d}\widetilde{T}\left(\varepsilon\right)\left[f_{R}-f_{L}\right]d\varepsilon.
\end{eqnarray*}

\subsection{Final expression of the curl}

The results of sections \ref{app:rho=E_d}, \ref{app:rho=V_L} and \ref{app:rho=V_R}, can be written in the compact form
\begin{eqnarray*}
\left(\nabla\times\overrightarrow{F}\right) & = & \stackrel[-\infty]{\infty}{\int}\overrightarrow{g}\left(E_{d},V_{L},V_{R}\right)N_{d}\widetilde{T}\left(\varepsilon\right)\left[f_{L}-f_{R}\right]d\varepsilon,
\end{eqnarray*}
where
\begin{eqnarray*}
\overrightarrow{g}\left(E_{d},V_{L},V_{R}\right) & = & \left(2\frac{\varepsilon-E_{d}}{V_{L}V_{R}},\frac{1}{V_{R}},\frac{1}{V_{L}}\right) .
\end{eqnarray*}

\section{Elements of the electronic-friction tensor in the TB system.\label{app:friction_TB}}
The three by three equilibrium friction tensor is symmetric, thus only six elements are necessary to describe it.
To derive the expression for the elements of $\boldsymbol{\gamma}$ we used the following relations
\[
\begin{array}{ccc}
\boldsymbol{\Lambda}_{V_{L}}\boldsymbol{A} & = & -2\begin{pmatrix}\textrm{Im}G_{12}^{r} & \textrm{Im}G_{22}^{r} & \textrm{Im}G_{23}^{r}\\
\textrm{Im}G_{11}^{r} & \textrm{Im}G_{12}^{r} & \textrm{Im}G_{13}^{r}\\
0 & 0 & 0
\end{pmatrix}\end{array} ,
\]
\[
\begin{array}{ccc}
\boldsymbol{\Lambda}_{V_{R}}\boldsymbol{A} & = & -2\begin{pmatrix}0 & 0 & 0\\
\textrm{Im}G_{13}^{r} & \textrm{Im}G_{23}^{r} & \textrm{Im}G_{33}^{r}\\
\textrm{Im}G_{12}^{r} & \textrm{Im}G_{22}^{r} & \textrm{Im}G_{23}^{r}
\end{pmatrix}\end{array}
\]
and
\[
\begin{array}{ccc}
\boldsymbol{\Lambda}_{E_{d}}\boldsymbol{A} & = & 2\begin{pmatrix}0 & 0 & 0\\
\textrm{Im}G_{12}^{r} & \textrm{Im}G_{22}^{r} & \textrm{Im}G_{23}^{r}\\
0 & 0 & 0
\end{pmatrix}\end{array}
\]
The above come from the definition of the spectral function, $\boldsymbol{A}=i\left(\boldsymbol{G}^{r}-\boldsymbol{G}^{a}\right)$,
and the matrices $\boldsymbol{\Lambda}_{\nu}$, Eq. \ref{eq:Lambda}.
To evaluate the elements of the above matrices one can use the explicit expressions for the elements of the retarded Green's function given in appendix \ref{app:GFs}.
After some algebra we find
\begin{eqnarray}
\textrm{Im}G_{11}^{r}&=&-\frac{\Gamma_{R}}{V_{0}^{2}}\frac{1}{\left|\varepsilon-E_{d}-\Sigma_{R}-\Sigma_{L}\right|^{2}}
\notag \\ & & 
\left\{ \left(\varepsilon-E_{d}\right)^{2}\left(\frac{V_{R}}{V_{0}}\right)^{-2}-2\left(\varepsilon-E_{d}\right)\Delta_{0}\right.
\notag \\ & &
\left.+\left[\left(\frac{V_{R}}{V_{0}}\right)^{2}+\left(\frac{V_{L}}{V_{0}}\right)^{2}\right]\left|\Sigma_{0}^{r}\right|^{2}\right\} \label{eq:ImG11}
\end{eqnarray}
\begin{eqnarray}
\textrm{Im}G_{22}^{r} & = & -\frac{1}{\left|\varepsilon-E_{d}-\Sigma_{R}-\Sigma_{L}\right|^{2}}\left[\Gamma_{L}+\Gamma_{R}\right]
\end{eqnarray}
\begin{eqnarray}
\textrm{Im}G_{12}^{r} & = & \frac{\left(\varepsilon-E_{d}\right)}{V_{L}}\frac{1}{\left|\varepsilon-E_{d}-\Sigma_{R}-\Sigma_{L}\right|^{2}}\Gamma_{L}
\end{eqnarray}
\begin{eqnarray}
\textrm{Im}G_{13}^{r}&=&\frac{1}{V_{L}V_{R}}\frac{\left(\frac{V_{L}}{V_{0}}\right)^{2}\left(\frac{V_{R}}{V_{0}}\right)^{2}\Gamma_{0}}{\left|\varepsilon-E_{d}-\Sigma_{R}-\Sigma_{L}\right|^{2}}
\left\{ 2\left(E_{d}-\varepsilon\right)^{ }_{ } \Delta_{0}\right. \notag \\
&&\left.+\left|\Sigma_{0}^{r}\right|^{2}\left[\left(\frac{V_{R}}{V_{0}}\right)^{2}+\left(\frac{V_{L}}{V_{0}}\right)^{2}\right]\right\}
\end{eqnarray}
\begin{eqnarray}
\textrm{Im}G_{23}^{r} & = & \frac{\left(\varepsilon-E_{d}\right)}{V_{R}}\frac{1}{\left|\varepsilon-E_{d}-\Sigma_{R}-\Sigma_{L}\right|^{2}}\Gamma_{R} \label{eq:ImG23}
\end{eqnarray}
Using the above formulas, and the expressions for the transmitance $\widetilde{T}\left(\varepsilon\right)$ and the local density of states $N_{d}\left(\varepsilon\right)$ (Eqs. \ref{eq:T_TB} and \ref{eq:Nd_TB} respectively), the elements of the electronic-friction tensor yield:
\subsection{Element $\gamma_{E_{d}E_{d}}^{\mathrm{eq}}$.}
\[
\begin{array}{ccc}
\underset{k_{B}T_{0}\rightarrow0^{+}}{\lim}\gamma_{E_{d}E_{d}}^{\mathrm{eq}} & = & \frac{1}{4\pi}\mathbf{\textrm{Tr}}\left[\Lambda_{E_{d}}A\Lambda_{E_{d}}A\right]_{\varepsilon=\mu_0}
\\
 & = & \pi N_{d}\left(\mu_0\right)N_{d}\left(\mu_0\right)
\end{array}
\]
where we used $\begin{array}{ccc}
\mathbf{\textrm{Tr}}\left[\Lambda_{E_{d}}A\Lambda_{E_{d}}A\right] & = & 4\textrm{Im}G_{22}^{r}\textrm{Im}G_{22}^{r}\end{array}$.
\subsection{Element $\gamma_{V_{L}V_{L}}^{\mathrm{eq}}$.}
\[
\begin{array}{ccc}
\underset{k_{B}T_{0}\rightarrow0^{+}}{\lim}\gamma_{V_{L}V_{L}}^{\mathrm{eq}} & = & \frac{1}{4\pi}\mathbf{\textrm{Tr}}\left[\Lambda_{V_{L}}A\Lambda_{V_{L}}A\right]_{\varepsilon=\mu_0} \\
 & = & 4\pi\frac{\left(\mu_0-E_{d}\right)^{2}V_{L}^{2}}{\left\{ V_{R}^{2}+V_{L}^{2}\right\} ^{2}}N_{d}\left(\mu_0\right)N_{d}\left(\mu_0\right)  \\ &&
 +\frac{1}{2\pi}\frac{1}{V_{L}^{2}} \widetilde{T}\left(\mu_0\right)
\end{array}
\]
Here we used 
\[
\begin{array}{ccc}
\mathbf{\textrm{Tr}}\left[\Lambda_{V_{L}}A\Lambda_{V_{L}}A\right] & = & 8\left(\textrm{Im}G_{12}^{r}\textrm{Im}G_{12}^{r}+\textrm{Im}G_{22}^{r}\textrm{Im}G_{11}^{r}\right)\end{array}
\]
and
\begin{eqnarray*}
\textrm{Im}G_{12}^{r}\textrm{Im}G_{12}^{r}+\textrm{Im}G_{22}^{r}\textrm{Im}G_{11}^{r}=&&\\2\pi^{2}\frac{\left(\varepsilon-E_{d}\right)^{2}V_{L}^{2}}{\left\{ V_{R}^{2}+V_{L}^{2}\right\} ^{2}}\left(-\frac{1}{\pi}\textrm{Im}G_{22}^{r}\right)\left(-\frac{1}{\pi}\textrm{Im}G_{22}^{r}\right)&&\\+\frac{1}{4V_{L}^{2}}4\frac{1}{\left|\varepsilon-E_{d}-\Sigma_{R}-\Sigma_{L}\right|^{2}}\Gamma_{R}\Gamma_{L}&&
\end{eqnarray*}
The latter expression requires some algebra but comes from Eqs. \ref{eq:ImG11}-\ref{eq:ImG23}.
\subsection{Element $\gamma_{V_{R}V_{R}}^{\mathrm{eq}}$.}
\[
\begin{array}{ccc}
\underset{k_{B}T_{0}\rightarrow0^{+}}{\lim}\gamma_{V_{R}V_{R}}^{\mathrm{eq}} & = & \frac{1}{4\pi}\mathbf{\textrm{Tr}}\left[\Lambda_{V_{R}}A\Lambda_{V_{R}}A\right]_{\varepsilon=\mu_0}\\
 & = & 4\pi\frac{\left(\mu_0-E_{d}\right)^{2}V_{R}^{2}}{\left\{ V_{R}^{2}+V_{L}^{2}\right\} ^{2}}N_{d}\left(\mu_0\right)N_{d}\left(\mu_0\right) \\ &&
 +\frac{1}{2\pi}\frac{1}{V_{R}^{2}}\widetilde{T}\left(\mu_0\right)
\end{array}
\]
Due to the symmetry of the problem, this element is equal to $\gamma_{V_{L}V_{L}}^{\mathrm{eq}}$ but replacing $V_{L}$ by $V_{R}$ in the formulas.
\subsection{Element $\gamma_{E_{p}V_{L}}^{\mathrm{eq}}$.}
\[
\begin{array}{ccc}
\underset{k_{B}T_{0}\rightarrow0^{+}}{\lim}\gamma_{E_{d}V_{L}}^{\mathrm{eq}} & = & \frac{1}{4\pi}\mathbf{\textrm{Tr}}\left[\Lambda_{E_{d}}A\Lambda_{V_{L}}A\right]_{\varepsilon=\mu_0}\\
 & = & 2\pi\frac{\left(\mu-E_{d}\right)V_{L}}{V_{L}^{2}+V_{R}^{2}}N_{d}\left(\mu_0\right)N_{d}\left(\mu_0\right)
\end{array}
\]
where we used $\begin{array}{ccc}
\mathbf{\textrm{Tr}}\left[\Lambda_{E_{d}}A\Lambda_{V_{L}}A\right] & = & -8\textrm{Im}G_{22}^{r}\textrm{Im}G_{12}^{r}\end{array}$.
\subsection{Element $\gamma_{E_{d}V_{R}}^{\mathrm{eq}}$.}
\[
\begin{array}{ccc}
\underset{k_{B}T_{0}\rightarrow0^{+}}{\lim}\gamma_{E_{d}V_{R}}^{\mathrm{eq}} & = & \frac{1}{4\pi}\mathbf{\textrm{Tr}}\left[\Lambda_{E_{P}}A\Lambda_{V_{R}}A\right]_{\varepsilon=\mu_0}\\
 & = & 2\pi\frac{\left(\mu-E_{d}\right)V_{R}}{V_{L}^{2}+V_{R}^{2}}N_{d}\left(\mu_0\right)N_{d}\left(\mu_0\right)
\end{array}
\]
Due to the symmetry of the problem, this element is equal to $\gamma_{E_{p}V_{L}}^{\mathrm{eq}}$
but replacing $V_{L}$ by $V_{R}$ in the formulas.
\subsection{Element $\gamma_{V_{L}V_{R}}^{\mathrm{eq}}$.}
\[
\begin{array}{ccc}
\underset{k_{B}T_{0}\rightarrow0^{+}}{\lim}\gamma_{V_{L}V_{R}}^{\mathrm{eq}} & = & \frac{1}{4\pi}\mathbf{\textrm{Tr}}\left[\Lambda_{V_{L}}A\Lambda_{V_{R}}A\right]_{\varepsilon=\mu_0}\\
 & = & 4\pi\frac{\left(\mu_0-E_{d}\right)^{2}V_{L}V_{R}}{\left\{ V_{R}^{2}+V_{L}^{2}\right\} ^{2}}N_{d}\left(\mu_0\right)N_{d}\left(\mu_0\right) \\ &&
 -\frac{1}{2\pi V_{L}}\frac{1}{V_{R}}\widetilde{T}\left(\mu_0\right)
\end{array}
\]
Here we used
\[
\begin{array}{ccc}
\mathbf{\textrm{Tr}}\left[\Lambda_{V_{L}}A\Lambda_{V_{R}}A\right] & = & 8\left(\textrm{Im}G_{22}^{r}\textrm{Im}G_{13}^{r}+\textrm{Im}G_{23}^{r}\textrm{Im}G_{12}^{r}\right)\end{array}
\]
and
\begin{eqnarray*}
\textrm{Im}G_{22}^{r}\textrm{Im}G_{13}^{r}+\textrm{Im}G_{23}^{r}\textrm{Im}G_{12}^{r}=&&\\2\pi^{2}\frac{\left(\varepsilon-E_{d}\right)^{2}V_{L}V_{R}}{\left\{ V_{R}^{2}+V_{L}^{2}\right\} ^{2}}\left(-\frac{1}{\pi}\textrm{Im}G_{22}^{r}\right)\left(-\frac{1}{\pi}\textrm{Im}G_{22}^{r}\right)&&\\-\frac{1}{4V_{L}}\frac{1}{V_{R}}4\frac{1}{\left|\varepsilon-E_{d}-\Sigma_{R}-\Sigma_{L}\right|^{2}}\Gamma_{L}\Gamma_{R}&&
\end{eqnarray*}
The latter expression requires some algebra but comes from Eqs. \ref{eq:ImG11}-\ref{eq:ImG23}.

\bibliography{cite-arxiv-01}

\end{document}